\documentclass[acmsmall, screen, nonacm]{acmart}

\usepackage{longtable}
\usepackage{capt-of}
\usepackage{balance,bm}
\usepackage{color, colortbl, xcolor}
\definecolor{LightGray}{gray}{0.97}
\usepackage{url}

\usepackage{subcaption}
\usepackage{booktabs} % For formal tables
\usepackage{graphicx}
\usepackage{textcomp}
\usepackage{color,soul}
\usepackage{bm}
\usepackage{multirow}
\usepackage{wrapfig}
\usepackage{balance}
\usepackage{graphicx}  %Required
\usepackage{enumitem}

\usepackage{colortbl}
\usepackage{arydshln}
\usepackage [english]{babel}
\usepackage [autostyle, english = american]{csquotes}
\definecolor{linkColor}{RGB}{6,125,233}
\definecolor{green}{rgb}{0.0, 0.65, 0.31}
\definecolor{bleudefrance}{rgb}{0.19, 0.55, 0.91}
\definecolor{ceruleanblue}{rgb}{0.16, 0.32, 0.75}
\definecolor{grey}{HTML}{969696}
\definecolor{violet}{HTML}{756bb1}
\definecolor{dgrey}{HTML}{01665e}
\definecolor{lgrey}{HTML}{5ab4ac}
\definecolor{dgreen}{HTML}{005a32}
\definecolor{purple}{HTML}{ae017e}

\definecolor{maskCol}{HTML}{c51b7d}
\definecolor{lrColor}{HTML}{8856a7}
\definecolor{trColor}{HTML}{d01c8b}
\definecolor{ctColor}{HTML}{4dac26}
\definecolor{brickred}{HTML}{f03b20}
\definecolor{improveCol}{HTML}{253494}
\definecolor{worsenCol}{HTML}{d7191c}
\definecolor{DarkBlue}{HTML}{00008B}
\definecolor{mscolor}{HTML}{01665e}
\definecolor{nmscolor}{HTML}{bf812d}
\definecolor{lgreen}{HTML}{ccece6}
\definecolor{dolive}{HTML}{308014}

\definecolor{maskCol}{HTML}{c51b7d}
\definecolor{lrColor}{HTML}{8856a7}
\definecolor{trColor}{HTML}{d01c8b}
\definecolor{ctColor}{HTML}{4dac26}
\definecolor{brickred}{HTML}{f03b20}
\definecolor{improveCol}{HTML}{253494}
\definecolor{worsenCol}{HTML}{d7191c}
\definecolor{lgreen}{HTML}{e0f3db}
\definecolor{dpink}{HTML}{CD1076}
\definecolor{pink}{HTML}{FED2D2}
\definecolor{soothinggreen}{HTML}{4dac26}
\definecolor{darkred}{HTML}{8B0000}

\definecolor{dblue}{HTML}{104E8B}
\definecolor{violet}{HTML}{8A2BE2}
\definecolor{mscolor}{HTML}{01665e}
\definecolor{nmscolor}{HTML}{d8b365}
\definecolor{deepgrey}{HTML}{525252}
\definecolor{dslate}{HTML}{2F4F4F}
\definecolor{dolive}{HTML}{556B2F}
\definecolor{teal}{HTML}{388E8E}
\definecolor{mscolor}{HTML}{01665e}
\definecolor{nmscolor}{HTML}{d8b365}

\definecolor{aicolor}{HTML}{018571}
\definecolor{occolor}{HTML}{ff7799}

\definecolor{srcolor}{HTML}{e34a33}
\definecolor{smcolor}{HTML}{253494}
\definecolor{srsmcolor}{HTML}{7fcdbb}
\definecolor{bothcolor}{HTML}{fe9929}
\definecolor{onecolor}{HTML}{018571}
\definecolor{marroon}{HTML}{881c1c}

\usepackage{mathtools}

\usepackage{amsmath}
\usepackage{array}
\usepackage{xcolor}
\usepackage{arydshln}
\usepackage{siunitx}
\colorlet{tablerowcolor4}{gray!50} % Table row separator colour = 

\newcommand*{\textlabel}[2]{%
  \edef\@currentlabel{#1}% Set target label
  \phantomsection% Correct hyper reference link
  #1\label{#2}% Print and store label
}
\usepackage{tcolorbox}

\colorlet{tableheadcolor}{gray!25} % Table header colour = 25% gray
\colorlet{tablerowcolor}{gray!15} % Table row separator colour = 
\colorlet{tablerowcolor2}{gray!45} % Table row separator colour = 
\colorlet{tablerowcolor3}{gray!25} % Table row separator colour = 10% gray

\newcommand{\rowcolmedium}{\rowcolor{tablerowcolor2}}
\newcommand{\rowcollight}{\rowcolor{LightGray}} %
\usepackage{capt-of}
\usepackage{balance,bm}
\usepackage{color, colortbl, xcolor}

\usepackage{tcolorbox}

\usepackage{booktabs}  
\usepackage{url}
\usepackage{hyperref}
\hypersetup{
    colorlinks=true,
}

\usepackage{subcaption}
\usepackage{booktabs} % For formal tables
\usepackage{graphicx}
\usepackage{textcomp}
\usepackage{color,soul}
\usepackage{bm}
\usepackage{multirow}
\usepackage{wrapfig}
\usepackage{balance}
\usepackage{graphicx}  %Required
\usepackage{enumitem}

\usepackage{colortbl}
\usepackage{arydshln}
\usepackage [english]{babel}
\usepackage [autostyle, english = american]{csquotes}
\definecolor{linkColor}{RGB}{6,125,233}
\definecolor{green}{rgb}{0.0, 0.65, 0.31}
\definecolor{bleudefrance}{rgb}{0.19, 0.55, 0.91}
\definecolor{ceruleanblue}{rgb}{0.16, 0.32, 0.75}
\definecolor{grey}{HTML}{969696}
\definecolor{violet}{HTML}{756bb1}
\definecolor{dgrey}{HTML}{01665e}
\definecolor{lgrey}{HTML}{5ab4ac}
\definecolor{dgreen}{HTML}{005a32}
\definecolor{purple}{HTML}{ae017e}

\definecolor{editCol}{HTML}{000000}
\definecolor{minorCol}{HTML}{000000}
\definecolor{maskCol}{HTML}{c51b7d}
\definecolor{lrColor}{HTML}{8856a7}
\definecolor{trColor}{HTML}{d01c8b}
\definecolor{ctColor}{HTML}{4dac26}
\definecolor{brickred}{HTML}{f03b20}
\definecolor{improveCol}{HTML}{253494}
\definecolor{worsenCol}{HTML}{d7191c}
\definecolor{DarkBlue}{HTML}{00008B}
\definecolor{mscolor}{HTML}{01665e}
\definecolor{nmscolor}{HTML}{bf812d}
\definecolor{lgreen}{HTML}{ccece6}
\definecolor{dolive}{HTML}{308014}

\definecolor{maskCol}{HTML}{c51b7d}
\definecolor{lrColor}{HTML}{8856a7}
\definecolor{trColor}{HTML}{d01c8b}
\definecolor{ctColor}{HTML}{4dac26}
\definecolor{brickred}{HTML}{f03b20}
\definecolor{improveCol}{HTML}{253494}
\definecolor{worsenCol}{HTML}{d7191c}
\definecolor{lgreen}{HTML}{e0f3db}
\definecolor{dpink}{HTML}{CD1076}
\definecolor{pink}{HTML}{FED2D2}
\definecolor{soothinggreen}{HTML}{4dac26}
\definecolor{darkred}{HTML}{8B0000}

\definecolor{dblue}{HTML}{104E8B}
\definecolor{violet}{HTML}{8A2BE2}
\definecolor{mscolor}{HTML}{01665e}
\definecolor{nmscolor}{HTML}{d8b365}
\definecolor{deepgrey}{HTML}{525252}
\definecolor{dslate}{HTML}{2F4F4F}
\definecolor{dolive}{HTML}{556B2F}
\definecolor{teal}{HTML}{388E8E}
\definecolor{mscolor}{HTML}{01665e}
\definecolor{nmscolor}{HTML}{d8b365}

\definecolor{aicolor}{HTML}{018571}
\definecolor{occolor}{HTML}{ff7799}

\definecolor{srcolor}{HTML}{e34a33}
\definecolor{smcolor}{HTML}{253494}
\definecolor{srsmcolor}{HTML}{7fcdbb}
\definecolor{bothcolor}{HTML}{fe9929}
\definecolor{onecolor}{HTML}{018571}
\definecolor{marroon}{HTML}{881c1c}

\usepackage{mathtools}

\usepackage{amsmath}
\usepackage{array}
\usepackage{xcolor}
\usepackage{arydshln}
\usepackage{siunitx}
\colorlet{tablerowcolor4}{gray!50} % Table row separator colour = 

\usepackage{tcolorbox}

\colorlet{tableheadcolor}{gray!25} % Table header colour = 25% gray
\colorlet{tablerowcolor}{gray!15} % Table row separator colour = 
\colorlet{tablerowcolor2}{gray!45} % Table row separator colour = 
\colorlet{tablerowcolor3}{gray!25} % Table row separator colour = 10% gray

\newif{\ifhidecomments}
  \hidecommentsfalse 
\ifhidecomments
    \newcommand{\keran}[1]{}
    \newcommand{\melissa}[1]{}
    \newcommand{\dongwhi}[1]{}
    \newcommand{\koustuv}[1]{}
    \newcommand{\ravi}[1]{}
\else
    \newcommand{\keran}[1]{\textbf{\small\sffamily{\textcolor{DarkBlue}{[#1 -- Keran]}}}}
    \newcommand{\melissa}[1]{\textbf{\small\sffamily{\textcolor{dolive}{[#1 -- Melissa]}}}}
    \newcommand{\ravi}[1]{\textbf{\small\sffamily{\textcolor{marroon}{[#1 -- Ravi]}}}}
    \newcommand{\dongwhi}[1]{\textbf{\small\sffamily{\textcolor{dpink}{[#1 -- Dong Whi]}}}}
    \newcommand{\koustuv}[1]{\textbf{\small\sffamily{\textcolor{brickred}{[#1 -- Koustuv]}}}}
  \fi
\newcommand{\edit}[1]{{\textcolor{editCol}{#1}}}
\newcommand{\minor}[1]{{\textcolor{minorCol}{#1}}}

\renewcommand{\textrightarrow}{$\rightarrow$}

\colorlet{tableheadcolor}{gray!25} % Table header colour = 25% gray
\colorlet{tablerowcolor}{gray!5} % Table row separator colour = 10% gray

\definecolor{neutralCol}{HTML}{dd1c77}
\definecolor{neutralGreen}{HTML}{31a354}
\definecolor{NewBlue}{HTML}{1879ba}
\definecolor{bleudefrance}{rgb}{0.19, 0.55, 0.91}  
\definecolor{AfTrColor}{HTML}{0868ac}  
\definecolor{BfTrColor}{HTML}{a8ddb5}  

\definecolor{AfCtColor}{HTML}{b10026}  
\definecolor{BfCtColor}{HTML}{fd8d3c}

\graphicspath{ {figures/} }

\newcommand{\para}[1]{\vspace{0.3em}\noindent\textbf{#1}~}

\AtBeginDocument{%
  \providecommand\BibTeX{{%
    \normalfont B\kern-0.5em{\scshape i\kern-0.25em b}\kern-0.8em\TeX}}}

\acmJournal{PACMHCI}
\begin{document}

\title[Exploring Mental Health Needs in Alzheimer's and Dementia Caregiving]{Balancing Caregiving and Self-Care: Exploring Mental Health Needs of Alzheimer's and Dementia Caregivers}

% \title{A Comparative Analysis of LLM-based Chatbots and Online Communities in Providing Health Related Support for Alzheimer's Disease}

%%
%% The "author" command and its associated commands are used to define
%% the authors and their affiliations.
%% Of note is the shared affiliation of the first two authors, and the
%% "authornote" and "authornotemark" commands
%% used to denote shared contribution to the research.
\author{Jiayue Melissa Shi}
\orcid{0009-0007-0624-2421}
\authornote{These authors contributed equally to this work.}
\affiliation{%
  \institution{University of Illinois Urbana-Champaign}
 \city{Urbana}
 \state{IL}
 \country{USA}}
 \email{mshi24@illinois.edu}

\author{Keran Wang}
\authornotemark[1]
\orcid{0009-0001-2462-2272}
\affiliation{%
  \institution{University of Illinois Urbana-Champaign}
 \city{Urbana}
 \state{IL}
 \country{USA}}
 \email{keranw2@illinois.edu}

\author{Dong Whi Yoo}
\orcid{0000-0003-2738-1096}
\affiliation{%
 \institution{Indiana University Indianapolis}
 \city{Indianapolis}
 \state{IN}
 \country{USA}}
 \email{dy22@iu.edu}

\author{Ravi Karkar}
\orcid{0000-0003-1467-4439}
\affiliation{%
 \institution{University of Massachusetts Amherst}
 \city{Amherst}
 \state{MA}
 \country{USA}}
 \email{rkarkar@umass.edu}

\author{Koustuv Saha}
\orcid{0000-0002-8872-2934}
\affiliation{%
 \institution{University of Illinois Urbana-Champaign}
 \city{Urbana}
 \state{IL}
 \country{USA}}
 \email{ksaha2@illinois.edu}

%%
%% By default, the full list of authors will be used in the page
%% headers. Often, this list is too long, and will overlap
%% other information printed in the page headers. This command allows
%% the author to define a more concise list
%% of authors' names for this purpose.
% \renewcommand{\shortauthors}{Jiayue Melissa Shi and Keran Wang et al.}

\renewcommand{\shortauthors}{Jiayue Melissa Shi et al.}

%%
%% By default, the full list of authors will be used in the page
%% headers. Often, this list is too long, and will overlap
%% other information printed in the page headers. This command allows
%% the author to define a more concise list
%% of authors' names for this purpose.

%%
%% The abstract is a short summary of the work to be presented in the
%% article.
%TC:ignore

\begin{abstract}
% \koustuv{Abstract needs to be updated with the new updates.}
Alzheimer's Disease and Related Dementias (AD/ADRD) are progressive neurodegenerative conditions that impair memory, thought processes, and functioning. 
Family caregivers of individuals with AD/ADRD face significant mental health challenges due to long-term caregiving responsibilities. 
Yet, current support systems often overlook the evolving nature of their mental wellbeing needs. 
Our study examines caregivers' mental wellbeing concerns, focusing on the practices they adopt to manage the burden of caregiving and the technologies they use for support. Through semi-structured interviews with 25 family caregivers of individuals with AD/ADRD, we identified the key causes and effects of mental health challenges \minor{and developed a temporal mapping of how caregivers' mental wellbeing evolves across three distinct stages of the caregiving journey.}
% and uncovered the evolution of caregivers' mental wellbeing over time. 
Additionally, our participants shared insights into improvements for existing mental health technologies, emphasizing the need for accessible, scalable, and personalized solutions \minor{that adapt to caregivers' changing needs over time.}
These findings offer a foundation for designing \minor{dynamic, stage-sensitive}
% tailored 
interventions that holistically support caregivers' mental wellbeing, benefiting both caregivers and care recipients. 

\end{abstract}
%TC:endignore

%%
%% The code below is generated by the tool at http://dl.acm.org/ccs.cfm.
%% Please copy and paste the code instead of the example below.
%%

% \ccsdesc[300]{Human-centered computing~Social media}

%%
%% Keywords. The author(s) should pick words that accurately describe
%% the work being presented. Separate the keywords with commas.

%%
%% This command processes the author and affiliation and title
%% information and builds the first part of the formatted document.
\maketitle

% \ifhidecomments
% \else
%     \thispagestyle{firststyle} % applies firststyle to first page
%     \pagestyle{allstyle}
% \fi

\section{Introduction}\label{section:intro}

About one in nine people (10.9\%) age 65 and older in the U.S. have Alzheimer’s Disease and Related Dementias (AD/ADRD)~\cite{rajan2021population}.
In 2023, 11.5M caregivers of people living with AD/ADRD provided an estimated 18.4 billion hours, or nearly 31 hours per week, of unpaid help~\cite{rabarison2018economic}.
Caregiving for AD/ADRD is an emotionally and physically demanding role, predominantly undertaken by informal and family caregivers at home~\cite{garcia2011anxiety,grunfeld1997caring}. 
% often filled by informal and family caregivers~\cite{garcia2011anxiety}. 
In fact, 74\% of AD/ADRD family caregivers are concerned about maintaining their own health since becoming a caregiver~\cite{Alz2024}.
They face a multitude of challenges throughout the caregiving journey, including managing the progressive symptoms of AD/ADRD, handling financial strains, and managing own wellbeing~\cite{unbehaun2018exploring,vaingankar2013perceived,wawrziczny2017spouse,gibson2014exploring,coon2009empirically}. 
Compared to formal caregivers, family caregivers face more pronounced challenges due to factors such as emotional attachment to the care recipient, lack of formal training, and insufficient resources and equipment~\cite{bull1990factors,duplantier2023barriers,reinhard2008supporting}.
As a result, they often struggle with high levels of stress, anxiety, and depression~\cite{pinquart2003associations, livingston1996depression,hughes2014correlates,cho2023effect}---at least one in three family caregivers of AD/ADRD individuals was found to suffer from clinical depression as per prior meta-analysis~\cite{sallim2015prevalence,ying2019validity}.

The mental health burden borne by family caregivers underscores the critical need for effective support systems and interventions to help them cope with the psychological demands of caregiving~\cite{coon2009empirically,harper2022}.
However, the specific mental wellbeing needs of AD/ADRD caregivers, and how these needs evolve throughout the caregiving journey, remain less understood. 
From a human-computer interaction (HCI) and computer-supported cooperative work (CSCW) perspective, there is also a gap in understanding the specific technology and collaboration needs of caregivers. 
% \koustuv{Bring up the collaborative nature of caregiving, and why CSCW should take care of the mental health/technology/collaborative nature.}
In particular, caregiving is inherently multi-layered and collaborative, involving direct interactions between caregivers and recipients while also requiring coordination with family members, healthcare providers, government agencies, community resources, and supportive technological tools~\cite{schulz2016family,robinson2021person}. 
% This collaborative dynamic extends to balancing the needs of the care recipient alongside managing their own wellbeing, both physically and mentally.
% Caregiving is inherently a multi-layered, collaborative process that not only involves the direct interaction between caregivers and care recipients but also requires coordination with family members, healthcare providers, government agencies, community resources, and even supportive technological tools~\cite{}. 
% The collaborative nature of caregiving is also demonstrated in the role of coordinating with these stakeholders as well as balancing the needs of the car recipient, simultaneously staying physically and mentally healthy.
% In this collaborative web, caregivers often take on the role of coordinating various stakeholders while balancing the needs of the care recipient and managing their own mental health. 
% Consequently, multi-dimensional responsibilities become a major source of psychological stress for caregivers. 
%Thus, adopting a Computer-Supported Cooperative Work (CSCW) perspective is crucial to more comprehensively understand and address the ways in which collaborative efforts among these parties can support caregivers’ mental wellbeing. By doing so, we can work towards developing more integrative support systems that ease the psychological burden on caregivers and foster a more resilient caregiving network.
Prior work in the HCI/CSCW space has explored technologies such as smartphone applications and virtual assistants to facilitate collaboration between caregivers and care recipients, supporting tasks like medication management and routine coordination~\cite{kim2024opportunities, meyerhoff2022meeting, shreve2016dementia,boessen2017online,kourtis2019digital,theisz2025exploring}.
However, the role of technology in supporting caregivers' mental health remains largely underexplored.

Together, given the global increase in AD/ADRD, the growing reliance on informal and family caregivers~\cite{nichols2022estimation,Alz2024}, and the promises of HCI and personal health informatics in supporting mental health~\cite{slovak2024hci,EPSTEIN_IMWUT_2020}, it is essential to develop a deeper understanding of the caregiving journey and how technology can better support family caregivers' mental wellbeing.
As people's needs evolve, so do their expectations of technology.
Recent work in personal informatics has highlighted the need to design tools that not only understand individuals' goals but also adapt as those goals evolve, providing better support~\cite{munson2020importance, schroeder2019examining, sefidgar2024migrainetracker}. 
As an AD/ADRD caregiver's needs evolve over time, from seeking information about the condition to learning how to care to discovering effective lifestyle adaptations, their expectations from technology also evolve. 
Adopting a personal health informatics perspective for technology design, the key open question is
% we seek to explore in this paper is 
\textit{how we can design mental wellbeing technologies that offer tailored and timely support to meet the unique needs of family caregivers?} 
% \edit{Unlike prior work that primarily focuses on either caregiving tasks or care recipient needs, our study examines the full spectrum of mental wellbeing concerns and their root causes, providing a comprehensive framework for understanding caregiver experiences across the temporal dimension of care.}
% about how do we build mental wellbeing technologies that can provide tailored support to the specific needs of family caregiveres?
% \koustuv{We have mostly highlighted theoretical gaps. Is there one practical gap we can add that will help enhance the motivation.}

% personal health informatics in supporting mental health~\cite{}, it is essential to gain a deeper understanding of the care journey and how technology can better support caregivers' mental wellbeing.
% it is essential to more comprehensively understand and address the ways in which collaborative efforts among multi-parties can support caregivers’ mental wellbeing. 
% it is crucial to focus on these unmet mental health needs.
% Towards addressing the above gaps, 
Accordingly, our research investigates the evolving mental wellbeing needs and concerns of AD/ADRD caregivers\footnote{Since our work focuses on family caregivers, the term 'caregivers' will refer to family caregivers unless otherwise noted.}.
%throughout their caregiving jo
Our work is guided by the following research questions (RQs):

\begin{enumerate}
    \item[\textbf{RQ1:}] \edit{What are the primary causes and effects of mental health challenges among AD/ADRD caregivers,} and how do these challenges evolve evolve throughout the caregiving journey?
    
    \item[\textbf{RQ2:}] What practices do AD/ADRD caregivers adopt \edit{in response to these evolving challenges?}

    \item[\textbf{RQ3:}] What technologies do AD/ADRD caregivers use to manage mental wellbeing, and what challenges do they face with current and emerging technologies?

\end{enumerate}

We conducted 25 semi-structured interviews with family caregivers of AD/ADRD individuals in the U.S., 
focusing on their routines, mental health challenges, coping strategies, and technology use.
Using inductive qualitative coding and thematic analysis~\cite{braun2019reflecting}, we identified key themes aligned with our RQs.
Participants reported multifaceted mental health concerns, which we categorized into causes and effects across three stages of evolving needs. While they recognized the importance of socializing and seeking professional support, many felt constrained by time and competing responsibilities. In this regard, they described both self-care practices and external support mechanisms they rely on.
They also discussed the role of technologies---from monitoring tools to social platforms and AI chatbots---in supporting their wellbeing. Caregivers expressed mixed views: some were optimistic about the personalization and timeliness of AI support, while others were skeptical of its accuracy and lack of human connection. 
Finally, we found the key barriers to technology use, including cost, accessibility, and lack of personalization in current mental health tools.

\edit{Our work is unique in its examination of the dynamic evolution of caregivers' mental health concerns throughout the caregiving journey, an aspect largely overlooked in prior work that typically focuses on point-in-time assessments of caregiver burden.} Our study builds on and contributes to the body of work of supporting caregiver needs in HCI and CSCW~\cite{bhat2023we,meyerhoff2022meeting,kim2024opportunities,siddiqui2023exploring,wong2024voice,foong2024designing}.
We situate our study with the social support behavioral code (SSBC)~\cite{suhr2004social} to understand the various types of support that family caregivers seek. 
Additionally, we draw upon the ethics of care framework~\cite{gilligan2014moral,tronto2020moral} to examine the relationships and dependencies caregivers must navigate to adequately address their mental wellbeing concerns amidst the challenges of caregiving.
\edit{Our work adds to the body of work by making the following key contributions:}
% We make several key contributions to the existing body of research on AD/ADRD caregiving and mental wellbeing, as listed below:}

\begin{itemize}

\item \edit{A thematic characterization of the multifaceted mental health needs and concerns of caregivers, categorizing them into cause-and-effect relationships.} 

\item \edit{A novel temporal framework that maps the evolution of mental wellbeing practices across three distinct stages of the caregiving journey.}

% \item \edit{An understanding of caregivers' perceptions of accessible, personalized, and human-centered technologies that adapt to their evolving needs throughout the caregiving journey.} 

\item \edit{A mapping of caregivers' perceived challenges with existing technologies and their proposed improvements---highlighting the need for accessible, adaptive, and human-centered technological solutions that evolve alongside the caregiving journey.}

% \koustuv{make it mapping of challenges and proposed improvements.}

\end{itemize}

% First, we provide thematic analysis~\cite{braun2019reflecting} characterizing the multifaceted mental health needs and concerns, categorizing them into cause-and-effect relationships. Second, we develop a novel temporal framework that maps the evolution of caregivers' mental wellbeing practices across three distinct stages of the caregiving journey. Third, we evaluate caregivers' perceptions of accessible, personalized, and human-centered technological solutions that can adapt to caregivers' changing needs across the caregiving trajectory. These 
\edit{The above contributions collectively advance both theoretical understanding and practical approaches to supporting caregivers' mental wellbeing.}
We also discuss the implications of this work in terms of policymaking, technology design, and ethics.
% Our study builds on and contributes to the body of work of supporting caregiver needs in HCI and CSCW~\cite{bhat2023we,meyerhoff2022meeting,kim2024opportunities,siddiqui2023exploring,wong2024voice,foong2024designing}.
% We situate our study with the social support behavioral code (SSBC)~\cite{suhr2004social} to understand the various types of support that family caregivers seek. 
% Additionally, we draw upon the ethics of care framework~\cite{gilligan2014moral,tronto2020moral} to examine the relationships and dependencies caregivers must navigate to adequately address their mental wellbeing concerns amidst the challenges of caregiving.
Our study offers empirical insights that lay a foundation for designing tailored support for family caregiver's mental wellbeing, ultimately benefiting both caregivers and care recipients. 
Our findings emphasize the collaborative nature of caregiving, which requires coordinated efforts among caregivers, care recipients, families, and communities. 
This dynamic highlights the importance of developing support systems that not only prioritize family caregivers' wellbeing but also enhance effective communication and resource-sharing within the caregiving network as well as the broader society.
% for support systems that not only address family caregiver wellbeing but also facilitate effective communication and resource-sharing across the caregiving network. 

 % \koustuv{need to add a summary of findings here once we have finalized our findings section.}
%We propose design opportunities for more personalized, scalable solutions, extending the current research on caregiver mental health and technological support.
 % \koustuv{Add how the discussion touches about the collaborative nature of caregiving.}

% \para{Privacy, Ethics, and Reflexivity.}
% % This paper used publicly accessible social media discussions on online communities and did not require any direct interactions with the individuals. 
% % Hence, it did not require an ethics board approval. 
% However, we are committed to the ethics of the research and followed practices to secure the privacy of the individuals in our dataset. 
% We recognize the sensitivities of our study in terms of revealing the identities of the individuals. 
% This paper only presents paraphrased quotes to reduce traceability yet provide context in readership. 
% Our research team comprises researchers holding diverse gender, racial, and cultural backgrounds, including people of color and immigrants, and hold interdisciplinary research expertise in the areas of HCI, CSCW, social computing, UbiComp, and Health Informatics. 
% \ravi{Koustuv, thoughts about commenting this last sentence to gain 2.5 lines back?}.\koustuv{gained the 2.5 lines.}
 %1.25 pages
% \input{2rw.tex} %1 page
\section{Background and Related Work}\label{section:rw}

% \koustuv{RW needs to change and be reorganized}

% \koustuv{RW needs to be updated with what is missing in prior work and what we do. }

\subsection{Alzheimer's Disease and Related Dementias (AD/ADRD): Condition and Caregiving}

Alzheimer's Disease and Related Dementias (AD/ADRD) is a family of 
neurodegenerative conditions that progressively worsens with no definitive \edit{cure}~\cite{kim2021scoping}. 
% progressive condition for which no definitive cure exists, and current treatments primarily focus on slowing its progression~\cite{kim2021scoping}. 
It remains a major public health concern, ranking as the fifth-leading cause of death among Americans aged 65 and older~\cite{Alz2024}. 
Projections indicate that by 2050, approximately 14M individuals in the U.S. and 152M worldwide will be living with AD/ADRD~\cite{nichols2022estimation}. 
The caregiving for AD/ADRD is predominantly undertaken by family members and informal caregivers within the home setting~\cite{garcia2011anxiety,grunfeld1997caring}.
In 2022 alone, the unpaid caregiving provided by family members was valued at approximately \$339.5B USD~\cite{Alz2024}.

Additionally, family caregivers frequently experience financial strain~\cite{gibson2014exploring,griffiths2016problems,unbehaun2018exploring} and difficulties in planning for future crises~\cite{millenaar2018exploring,vaingankar2013perceived,wawrziczny2017spouse}. 
Many also struggle with maintaining their own physical and mental wellbeing~\cite{coon2009empirically,hughes2014correlates,wawrziczny2017spouse}.
Therefore, this caregiving effort comes with significant personal costs, including a heightened risk of emotional distress and adverse mental and physical health outcomes for caregivers~\cite{harper2022,garcia2011anxiety,grunfeld1997caring,Alz2024}. 
% Consequently, the burden on caregivers, particularly informal and family caregivers, is significant~\cite{garcia2011anxiety,grunfeld1997caring,better2023alzheimer}. 
In particular, caregivers of AD/ADRD individuals often face higher levels of stress compared to those caring for individuals with other conditions~\cite{alzheimer2005alzheimer}.
Without sufficient training or support, they encounter numerous challenges, such as managing the evolving symptoms of the care recipient~\cite{rettinger2020mixed,ruitenburg2024evolving}, providing supervision~\cite{queluz2020understanding}, and making complex medical decisions regarding comorbid conditions~\cite{lee2022unmet}.

% \koustuv{@melissa, add the citations from Introduction in the above two paragraphs.}

% \minor{Our work extends this by examining how these mental health concerns evolve throughout the caregiving journey-a temporal dimension that remains underexplored in current literature.}
We build upon the prior work, highlighting the challenges faced by AD/ADRD caregivers, to understand the specific mental wellbeing concerns experienced in the AD/ADRD caregiving journey.
Our work unpacks the causes and effects of mental health concerns among the caregivers. \minor{Our work extends this by examining how these mental health evolve throughout the caregiving journey---a temporal dimension that remains underexplored in current literature.}
This helps us focus on the evolving mental wellbeing needs of these caregivers, and the barriers they face in navigating through the caregiving challenges and managing their own wellbeing.
% , which have not been sufficiently addressed in existing research. 
% By investigating how these needs change throughout the caregiving journey, we aim to offer a deeper understanding of the mental health challenges faced by caregivers. This focus goes beyond current studies and 
This study contributes to a more nuanced understanding of how caregiver support systems and technologies can be better tailored to address the dynamic and multifaceted concerns of family caregivers.

\subsection{Mental Health Needs of Informal and Family Caregivers}

Caregiving roles can include both professional and informal caregivers, our study specifically focuses on informal and family caregivers. 
Family caregivers are unpaid individuals, often family members or close acquaintances, who assist those with chronic or acute conditions by performing tasks ranging from daily care to complex medical procedures~\cite{reinhard2008supporting}. 
Due to the intensity and magnitude of responsibilities, family caregivers frequently face emotional overwhelm, which can lead to self-neglect and mental health issues such as depression~\cite{blum2010family}.

The prevalence of mental health issues among caregivers is a significant concern. 
Research reveals that caregivers often experience high levels of stress, depression, anxiety, and other psychological challenges due to the demands of caregiving~\cite{schulz2008physical}. 
Prior work indicates that many caregivers experience significant psychological stress, with 34.0\% reporting depression, 43.6\% facing anxiety, and 27.2\% resorting to psychotropic medications~\cite{sallim2015prevalence}. 
% These findings underscore the considerable mental health burden on caregivers and highlight the need for targeted support and interventions.

% Caregivers of individuals with Alzheimer's Disease (AD) face significant mental health challenges, often arising from their cognitive and emotional responses to the disease. 
Family caregivers frequently encounter stigma, as societal misconceptions about caregiving roles can lead to judgment and disapproval, exacerbating feelings of isolation and stress~\cite{corrigan2006blame,siddiqui2023exploring}.
% In the case of AD/ADRD, t
The progressive decline in the care recipient's memory and cognitive abilities, can provoke feelings of shame, embarrassment, and even disgust in caregivers~\cite{werner2010subjective}. 
% For example, caregivers may feel humiliated by socially inappropriate behavior exhibited by patients in public, or embarrassed by acts that others may witness.
This is especially problematic in contexts where community plays a crucial role, such as AD/ADRD care, where caregivers often desire recognition for their efforts beyond being seen as mere assistants~\cite{bhat2023we,siddiqui2023exploring,bosch2019caregiver}.
% Despite these negative emotions, 
Further, caregivers also experience compassion, sorrow, and guilt, driven by their deep desire to ease the suffering of their loved ones, coupled with grief over the person they feel they have lost~\cite{werner2010subjective}. 
However, this ongoing emotional strain can lead to compassion fatigue, causing caregivers to become emotionally drained and detached, making it more difficult for them to continue providing care---a phenomenon described as ``compassion fatigue''~\cite{day2011compassion}.
% The combination of conflicting emotions and compassion fatigue creates a heavy emotional burden that significantly impacts caregivers' mental health and overall wellbeing.

Further, the emotional toll on caregivers also bears repercussions for the care recipients~\cite{isik2019bidirectional}.
~\citeauthor{sun2022comparative} found that caregiver depression can accelerate cognitive decline in AD/ADRD individuals~\cite{sun2022comparative}. 
The deteriorating health of caregivers intensifies their caregiving burden, contributing to the worsening of the care recipient's condition~\cite{cho2023effect}.
% , their caregiving burden intensifies, which in turn contributes to the worsening of the patient's symptoms~\cite{cho2023effect}. 
Therefore, \citeauthor{del2019association} highlighted the importance of enhancing caregivers' perceived health~\cite{del2019association}.
% Thus, providing support to enhance caregivers' perceived health is vital not only to lessen their burden but also to improve patient health~\cite{del2019association}.
We build on the above body of work to examine AD/ADRD caregivers' mental health challenges, as well as their self-care practices, support mechanisms, and barriers such as insufficient support, compassion fatigue, and burnout. 
\edit{While prior research has highlighted caregiver mental health challenges, such as stress and compassion fatigue, these issues remain largely underexplored in CSCW and HCI. Caregivers increasingly interact with sociotechnical systems for support, making it vital for HCI/CSCW to understand their lived experiences and design more supportive technologies. Our work extends this literature by examining the evolving mental health needs of AD/ADRD caregivers and emphasizing the role of sociotechnical solutions in supporting their wellbeing.}

\subsection{HCI and CSCW Technologies for Caregivers' Wellbeing}

% The current landscape of wellbeing support for caregivers is shaped by a variety of technological tools and platforms
A rich body of prior work in HCI and CSCW has explored technologies to support caregivers both generally~\cite{bosch2019caregiver,miller2016partners,chen2013caring,seo2019balancing,jacobs2019think,lee2023reimagining,schorch2016designing,zubatiy2021empowering}, as well as for AD/ADRD~\cite{berridge2022advance,piper2016technological,stowell2019caring,carrasco2020empowering,lin2020go,smriti2024emotion,houben2024design}.
Prior work highlighted the benefits of wellbeing technologies in supporting AD/ADRD  caregivers~\cite{meiland2017technologies,yoon2020mining}. 
These include the use of smartphone~\cite{shreve2016dementia} and wearable~\cite{kourtis2019digital,stavropoulos2021wearable} technologies in supporting caregiver wellbeing.
In addition, research has explored the ethics and privacy surrounding the development and use of technologies to assist caregiving~\cite{Mulvenna2017,hodge2020relational,kropczynski2021examining,mentis2020illusion}.
Prior work explored how smartphone apps can help caregivers maintain routines and reduce stress, with features tailored to provide timely support~\cite{dayer2013smartphone}, and the use of Voice Interactive Personal Assistants (VIPAs) in providing mental health services to caregivers~\cite{wong2024voice,o2020voice,pradhan2020use}.
These technologies are especially useful for caregivers as they offer accessible, scalable support tailored to individual needs, ensuring mental health services are available when needed~\cite{randall2018engaging,lederman2019support}.

Further, online platforms provide a comprehensive array of resources aimed at improving caregiver mental health~\cite{johnson2022s,levonian2021patterns,johnson2022s,kaliappan2025online,saha2025ai}. These platforms often include telehealth services, monitoring tools, and self-care resources, allowing caregivers to manage stress and connect with professionals and peers~\cite{boessen2017online}. 
% By serving as virtual hubs, t
These platforms promote resilience and provide centralized, accessible support for caregivers in managing their mental health~\cite{boessen2017online}.
Over the last decade, chatbots have been integrated into healthcare, including AD/ADRD care, where they offer caregivers information, education, and support~\cite{ruggiano2021chatbots}. 
Despite the potential, chatbots for AD/ADRD caregiving is still in its early stages and requires more development to meet caregivers' needs effectively~\cite{ruggiano2021chatbots,wong2024voice,bosco2024don,bosco2025designing}.

Relatedly,~\citeauthor{bhat2023we} highlighted the critical mediating role played by caregivers, proposed caregiver-centric technological supports~\cite{bhat2023we}.
Likewise,~\citeauthor{kim2024opportunities} emphasized the role of customized support strategies, as caregiver needs fluctuate across different stages of challenging behavioral episodes~\cite{kim2024opportunities}, and \citeauthor{meyerhoff2022meeting} underscored the critical need for user-centered digital mental health (DMH) tools that adapt flexibly to individual support needs, which can empower users in their mental health journeys~\cite{meyerhoff2022meeting}. 
Recently,~\citeauthor{smriti2024emotion} explored if and how technologies can support caregivers of people living with dementia~\cite{smriti2024emotion}.

Parallelly, within HCI and CSCW, the personal informatics community has studied the design of technology to support mental health through a range of topics, including monitoring mood, affect, and stress to managing bipolar disorder and depression~\cite{EPSTEIN_IMWUT_2020, xu2023technology, blair2023knowing, burgess2022just,karusala2021future,yoo2025ai,saha2021job}. These studies explored how technology can be used to understand and manage mental health more effectively, highlighting the importance of tailored design in mental health technology~\cite{yoo2024patient,yoo2024missed}. 
% \koustuv{this blurb needs to be expanded with 2-3 sentences.} 

\edit{Although individual mental health has been extensively studied, our understanding of the mental health needs of informal and family caregivers---\textit{individuals whose mental health challenges arise not from their own condition, but from their ongoing responsibility for someone else's medical condition}---remains limited. This distinction is crucial, as caregiving introduces unique stressors that are not well represented in broader mental health research. Our study expands the scope of CSCW research to include the nuanced and evolving mental health needs of caregivers, whose experiences demand new forms of technological considerations.} 
% \minor{Unlike prior work that focuses on static technology solutions, o
\minor{We examine how AD/ADRD caregivers' mental health needs---and the technologies that support them---evolve across different stages of caregiving.}

By synthesizing two streams of research---addressing needs of people with mental health conditions and supporting caregivers in their care work---we aim to empirically inform the opportunities to address the specific mental wellbeing needs of caregivers.
Our work is further motivated by \citeauthor{chen2013caring}'s call for system design to focus on caregivers
% , addressing the burdens that impact their mental wellbeing 
and \citeauthor{kokorelias2022grounded} emphasis on the evolving needs of caregiving through different phases~\cite{chen2013caring,kokorelias2022grounded}.
We aim to understand caregivers' concerns and desires for technologies to address their caregiving and mental wellbeing challenges. 
Towards this aim, we explore the dynamic and multifaceted mental health needs, and their current practices to address mental health concerns.
Our work highlights ways to enhance technologies to better support caregivers' mental health throughout their caregiving journey, expanding the discourse on solutions in AD/ADRD caregiving.
 %1 page

\section{Study and Methods}
We conducted semi-structured interviews with caregivers of individuals with AD/ADRD.
We describe our methodology and participant pool in this section. 
% \edit{Our approach involved thematic analysis~\cite{braun2019reflecting} to identify patterns across participant experiences, with all participant quotes clearly labeled by ID throughout our findings. We selected our measures based on established frameworks in caregiving and mental health research, which we detail in the following sections.}
% , incorporating an in-interview survey to gain caregiving-related insights. 
% Our study was approved by the respective Institutional Review Boards (IRBs) at the researchers' institutions.
% We explain our methodology in the following subsections. 
% \subsection{Methodological Overview}
% \koustuv{We first synthesize taxonomies driven by the literature. Then, we will conduct an interview study by recruiting family caregivers from online communities. We will conduct inductive coding followed by thematic analyses to obtain themes from our interviews. Then, we will triangulate our observations from the literature and that from the interviews.}

% \subsection{Interview Study}

\subsection{Participants and Recruitment}
% \subsubsection{Recruitment and Participants}
We recruited our participants primarily through social media. 
We first contacted the moderators of different online communities catering to AD/ADRD-related discussions on Reddit (\textit{r/alzheimers, r/dementia, r/dementiaresearch, r/ParentsWithAlzheimers}, etc.) and AlzConnected (\textit{alzconnected.org}), by briefly describing our research and if they were okay with recruiting from their respective platforms.
Then, in online communities, where we received moderators' approval, we posted our recruitment flyer with an interest form that included a demographic survey questionnaire (age, sex, race, U.S. state) and their role as a caregiver. 
This interest form helped us target and screen participants who are 1) 18 years or older, 2) current/former caregivers for AD/ADRD, and 3) residing in the U.S.
We received 293 responses to our interest form over a period of two months between August and October 2024, and we invited a subset of participants to maximize diversity and balance across answers to the caregiving role and tenure.
This led to a final set of 25 participants who consented to participate in the study, and we interviewed them. 
We note that although our recruitment flyers were posted on social media, three of our participants did not actively participate on these platforms, and were rather referred to by others to express their interest in participating in our study.
Each participant was compensated with Amazon gift vouchers of \$25 USD.

\autoref{tab:participants} summarizes the demographic and caregiving role and tenure of participants. 
% We find that our participant pool includes individuals from a variety of age groups, 
% on pl AD/ADRD on Reddit, AlzConnected, 
% posted our recruitment flyers on a number of Reddit communities, such as \textit{r/alzheimers, r/dementia}, and {r/dementia}. 
% We interviewed 25 participants, all located in the U.S., with the following age distribution: 25-35 years: 13, 50-65 years: 8, 36-50 years: 2, and 66+ years: 2. 
% \autoref{tab:participants} summarizes the demographic and care recipient information of the participants. 
Of the participants we interviewed, 76\% (19 out of 25) are current caregivers, and we note a diversity of participants across age group, number of years in caregiving, race, education, and occupation.
Before the interviews, participants were provided with the Rapid Caregiver Well-being Scale (R-CWBS)~\cite{tebb2013caregiver} along with the consent form. 
R-CWBS is a validated short-form rapid assessment instrument to infer key areas of support a caregiver needs~\cite{tebb2013caregiver}.
Here, each question is rated on a Likert-scale between 1 (Rarely) and 5 (Usually), and lower scores indicate a need for greater support. 
\autoref{table:cwbs} provides a summary of participants' responses to this survey, showing that although our pool of participants was mostly regular with taking care of personal daily activities, the other questions received a variety of responses. 

% Within our participant pool, we found that 

% \autoref{table:cwbs} shows the mean responses for each item on the Rapid Caregiver Well-being Scale (R-CWBS)~\cite{tebb2013caregiver}, indicating how caregivers prioritize their lives. Using a numerical mapping (Usually = 5, Frequently = 4, Sometimes = 3, Occasionally = 2, Rarely = 1), we calculated the mean scores, with higher values for "Taking care of personal daily activities" (4.46) and "Receiving appropriate health care" (3.83) reflecting strong engagement in essential tasks. Lower scores in areas like "Treating or rewarding yourself" (3.21) suggest a need for more support in self-reward activities.

% Along with the consent fr
% , all of whom were meeting online at the time of the interview.
% We recruited caregivers through Alzheimer's community organizations. In an attempt to reach a wider range of caregivers, we also shared recruitment flyers on social media platforms such as Reddit. To be eligible for the study, participants needed to self-identify as current or former caregivers for AD/ADRD and be 18 years or older. 

\begin{table}[t!]
\centering
\sffamily
\footnotesize
   \caption{Demographic information regarding the participants, including Type (Current/Former Caregiver), Years of Caregiving (Ys.), Care Recipient, Age, Gender, Race, Education, and Occupation. Some caregivers started their caregiving for a family member, and later transitioned into professional caregiving. \textit{Serial Caregivers} are marked with an '*' next to their ID.}
   % \label{tab:pp1}
   \label{tab:participants}
\setlength{\tabcolsep}{2pt}
% \begin{tabular}{ccrcrcccc}
\resizebox{\columnwidth}{!}{
\begin{tabular}{llrlrllll}
\textbf{ID} & \textbf{Type} & \textbf{Ys.} & \textbf{Care Recipient} & \textbf{Age} & \textbf{Sex} & \textbf{Race} & \textbf{Education} & \textbf{Occupation} \\
\toprule
P1 & Current & 5 & Friend & 25-35 & Male & Asian & Associate degree& Employed for wages \\
\rowcollight P2* & Former  & 14 & Mother, Grandmothers & 50-65 & Female & White & Bachelor's& Retired \\
P3 & Current & 2 & Mother & 25-35 & Female & Black/African American & Associate degree& Out of work \\
\rowcollight P4 & Current & 1 & Father-in-law & 50-65 & Female & White & Some college, no deg. & Self-employed \\
P5 & Former  & 3 & Uncle & 50-65 & Female & White & Advanced degree& Employed for wages \\
\rowcollight P6 & Former  & 1 & Aunt & 25-35 & Male & White & Bachelor's& Employed for wages \\
P7 & Current & 1 & Grandfather & 25-35 & Male & Black/African American & Associate degree& Self-employed \\
\rowcollight P8 & Current & 6 & Grandfather & 25-35 & Male & Black/African American & Bachelor's& Employed for wages \\
P9 & Current  & 5 & Husband & 66+ & Female & White & Advanced degree & Retired \\
\rowcollight P10 & Former  & 4 & Mother & 36-50 & Female & White & Advanced degree& Employed for wages \\
P11 & Former  & .75 & Mother & 50-65 & Female & White & Advanced degree & Retired \\
\rowcollight P12 & Current  & 10 & Brother & 36-50 & Male & Black/African American & Advanced degree& Employed for wages \\
P13 & Current   & 1 & Wife & 66+ & Male & White & Bachelor's& Retired \\
\rowcollight P14 & Former  & 3 & Aunt & 25-35 & Female & Asian & Bachelor's& Self-employed \\
P15 & Current  & 1 & Family & 25-35 & Male & Black/African American & Some college, no deg. & Employed for wages \\
\rowcollight P16* & Current  & 5 & Grandfather, Clients & 25-35 & Female & Black/African American & Trade/technical & Employed for wages \\
P17* & Current  & 5 & Mother, Clients & 25-35 & Female & Black/African American & Trade/technical & Employed for wages \\
\rowcollight P18 & Current  & 5 & Father & 50-65 & Female & White & Bachelor's  & Employed for wages \\
P19 & Current & 7 & Mother & 50-65 & Female & Hispanic/Latino & Advanced degree & Employed for wages \\
\rowcollight P20 & Current & 6 & Mother & 25-35 & Female & Black/African American & Advanced degree & Employed for wages \\
P21* & Current & 6 & Mother, Clients   & 25-35 & Male & Black/African American & Trade/technical & Employed for wages \\
\rowcollight P22* & Current & 3 & Grandfather, Clients  & 25-35 & Female & Black/African American & Trade/technical & Employed for wages \\
P23 & Current & 5 & Mother  & 25-35 & Male & Black/African American & Trade/technical & Employed for wages \\
\rowcollight P24 & Current  & 5 & Mother  & 50-65 & Male & White & Bachelor's & Employed for wages \\
P25 & Current  & 10 & Husband  & 50-65 & Female & White & Advanced degree & Employed for wages \\
\bottomrule
\end{tabular}}
\Description[table]{}
\end{table}

\subsection{Interview Procedure}
% \koustuv{also include the caregivers wellbeing-scale and cite it that it was conducted along with the consent form.}

% \koustuv{We need to explain that during the interviews, we sought participants' experiences in the caregiving journey, and how it has impacted their mental health. We asked them about their daily caregiving routine,..}

We conducted semi-structured interviews with caregivers to explore their experiences and mental health throughout the caregiving journey.
We conducted these interviews via video calls (Zoom/Teams).
% , with participants receiving clear instructions on how to join the call. 
The research teams took turns interviewing and note-taking during the interviews. 
These interviews were recorded and lasted 60 minutes.
% We used \textit{otter.ai} for automated transcription of the five interviews conducted on Zoom, and used the default transcription on Teams for the remaining interviews.

% \edit{Before conducting the interviews, participants were provided with the Rapid Caregiver Wellbeing Scale (R-CWBS)~\cite{tebb2013caregiver} along with the consent form to assess their wellbeing status. This validated instrument helped us better understand participants' baseline wellbeing and contextualize their interview responses.}

% and how caregiving has impacted their mental health. 
During the interviews, we sought to understand participants' daily caregiving routines and how these responsibilities affected their mental wellbeing. 
% We discussed their experiences with mental health challenges and how their wellbeing evolved throughout the caregiving journey. 
% Additionally, w
We also inquired about the strategies they used to manage stress and cope with the mental health demands.
% emotional demands of caregiving.
% \edit{We deliberately chose not to collect household income data during our interviews due to privacy concerns and the sensitive nature of financial information in the caregiving context. Since many caregivers already experience financial strain, as revealed in our findings, we wanted to minimize any discomfort or potential barriers to participation.} 

\edit{Our interview protocol was systematically developed based on prior HCI/CSCW qualitative research with caregivers~\cite{johnson2022s,bhat2023we,kim2024opportunities,smriti2024emotion}. We began with open-ended questions about caregiving experiences to allow themes to emerge naturally, followed by more structured prompts derived from established literature.}
To gain further insight, we drew on the prior literature on the mental health concerns of AD/ADRD caregivers, and incorporated prompt questions.

We shared our screen with participants and asked them to rate eight caregiving-related concerns drawn from prior literature---1) disruptive patient behaviors~\cite{swearer1988troublesome,kunik2003disruptive,desai2012behavioral}, 2) insufficient support systems~\cite{schulz2004family, jennings2015unmet,park2018roles}, 3) doubt in self-efficacy~\cite{tang2016effects,gallagher2011self,crellin2014self}, 4) emotional wellness issue~\cite{aminzadeh2007emotional,caputo2021emotional,gallego2022feel}, 5) relationship management~\cite{ray2022impact,vroman2019impact}, 6) compassion fatigue~\cite{day2011compassion,day2014compassion,perez2022mindfulness}, 7) no time for self-care~\cite{waligora2019self,oliveira2019improving,wang2019impact}, and 8) burnout~\cite{alves2019burnout,truzzi2012burnout,takai2009experience}. 
% \edit{These eight concerns were obtained from prior literature as significant challenges for AD/ADRD caregivers. 
% \edit{This approach enabled us to situate our findings with existing knowledge while also allowing for the discovery of novel insights specific to our participants' experiences.}
We guided participants through each concern, providing literature-based definitions, and asked them to rate their level of concern on a scale from 1 (not at all concerning) to 5 (very concerning). 
Additionally, we asked the participants to explain their ratings and \textit{think aloud} about specific experiences related to these concerns. 
This helped establish shared understanding and yielded deeper insights into participants' mental wellbeing concerns which we qualitatively analyzed~\cite{braun2019reflecting}.
% This approach helped build common ground and helped us gather deeper insights into participants' mental wellbeing concerns.
\autoref{table:mentalwellbeing} summarizes participants' responses to these prompts, where we find a high concern for most of the prompts. 
% In our ensuing qualitative analyses~\cite{braun2019reflecting}, we incorporated the deeper insights that they shared. 
Finally, we asked the participants about the use of technology in caregiving as well as in managing mental wellbeing. 
Participants were encouraged to share their thoughts on the technologies they used, what features they found useful, and any suggestions or concerns they had for improving their experience. 
This helped us understand participants' concerns and desires about these technologies.
\begin{table}[t!]
\centering
\sffamily
\footnotesize
% \caption{Mean Responses for Each Question (Numerical)}
\caption{Summary of participants' responses to Rapid-Caregivers' Well-being Scale (R-CWBS)~\cite{tebb2013caregiver}. Each questions were rated on: 1 (Rarely), 2 (Occassionally), 3 (Sometimes), 4 (Frequently), and 5 (Usually).}
\label{table:cwbs}

\begin{tabular}{lrrrc}
\textbf{Question} & \textbf{Mean} & \textbf{\edit{Median}} & \textbf{Std. Dev.} & \textbf{Distribution}\\
\toprule
\rowcollight \multicolumn{4}{c}{\textbf{Activities}}\\
Taking care of personal daily activities (meals, hygiene, laundry) & 4.44 & \edit{4} & 0.57 & \includegraphics[height=6pt]{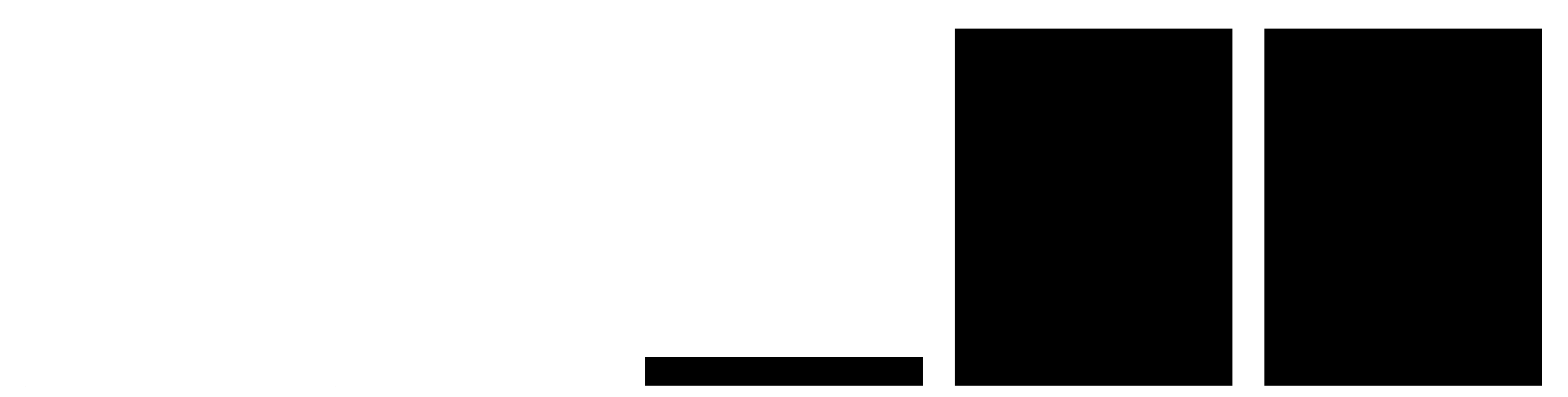}\\
Taking time to have fun with friends and/or family & 3.28 & \edit{4} & 1.25 & \includegraphics[height=6pt]{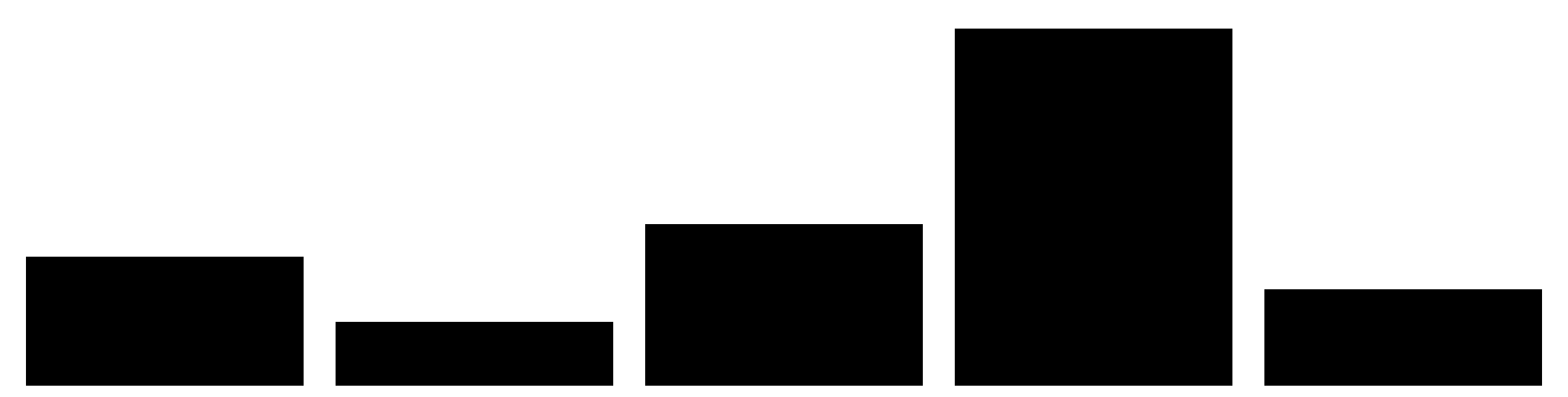}\\
Treating or rewarding yourself & 3.20 & \edit{3} & 1.23 & \includegraphics[height=6pt]{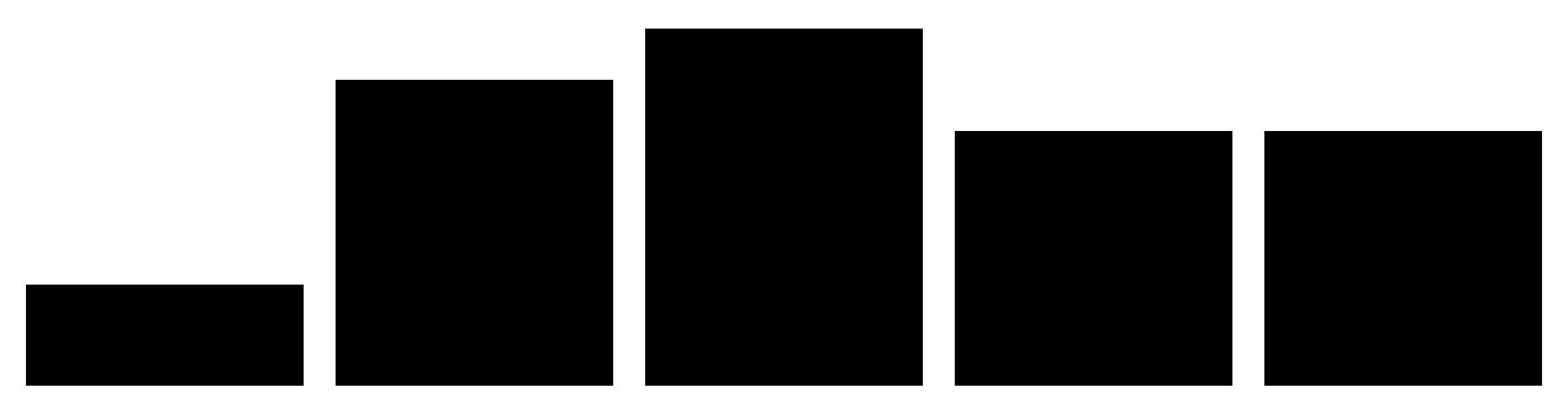}\\
\rowcollight \multicolumn{4}{c}{\textbf{Needs}}\\
Receiving appropriate health care & 3.84 & \edit{4} & 0.97 & \includegraphics[height=6pt]{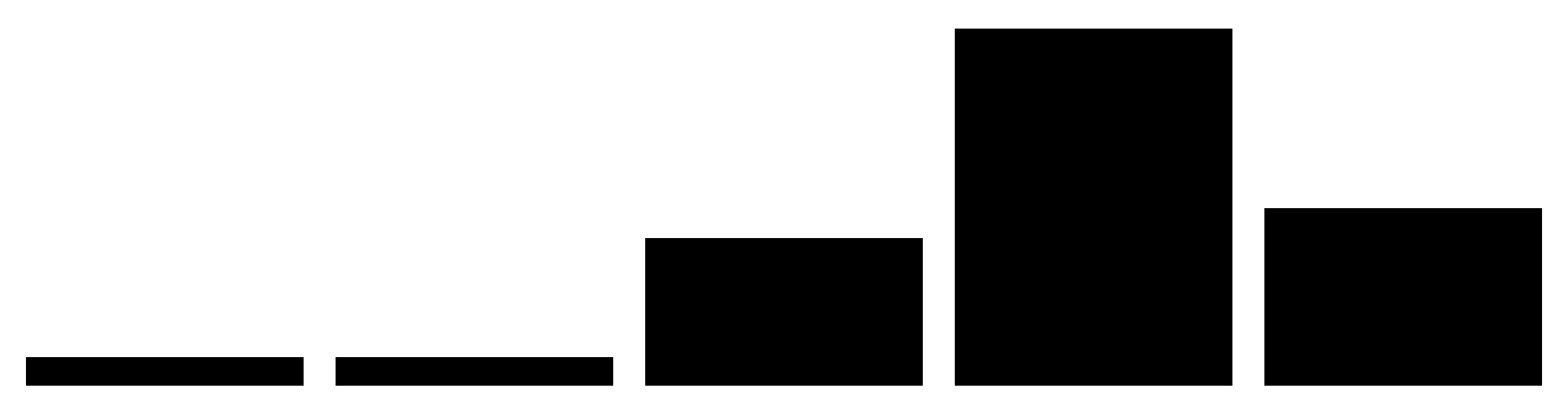}\\
Feeling good about yourself & 3.50 & \edit{4} & 1.20 & \includegraphics[height=6pt]{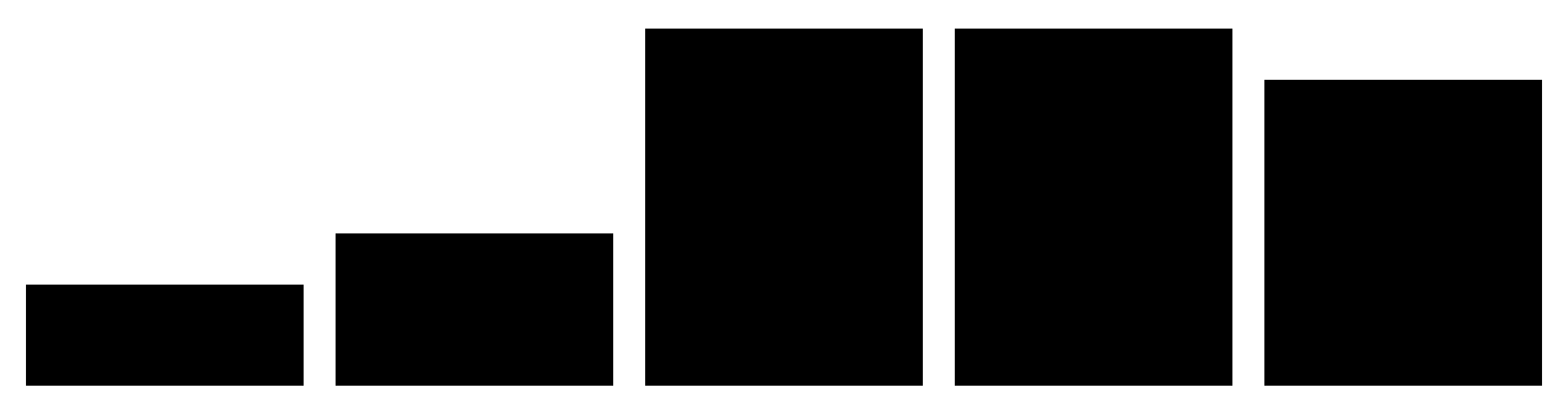}\\
Feeling secure about your financial future & 3.40 & \edit{3} & 1.33 & \includegraphics[height=6pt]{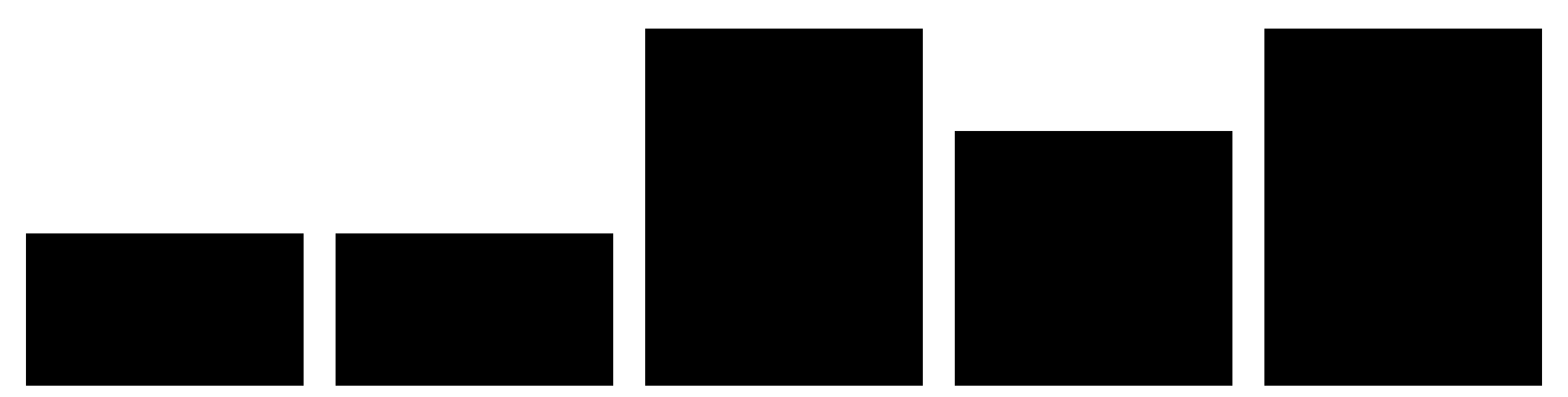}\\
\bottomrule
\end{tabular}

\end{table}

\begin{table}[t!]
\centering
\sffamily
\footnotesize
% \caption{Mean Responses for Each Question (Numerical)}
\caption{Summary of participants' responses to prompts on mental wellbeing concerns drawn on the literature. Participants responded to these prompts based on how much they associated with these concerns, on a scale of 1 (not at all concerning) to 5 (very concerning).}
\label{table:mentalwellbeing}
\begin{tabular}{lrrrc}
\textbf{Question} & \textbf{Mean} & \textbf{\edit{Median}} & \textbf{Std. Dev.} & \textbf{Distribution}\\
\toprule
% \rowcollight \multicolumn{4}{c}{\textbf{Activities}}\\
Disruptive Behaviors by Care Recipients & 4.00 & \edit{4} & 1.23 & \includegraphics[height=8pt]{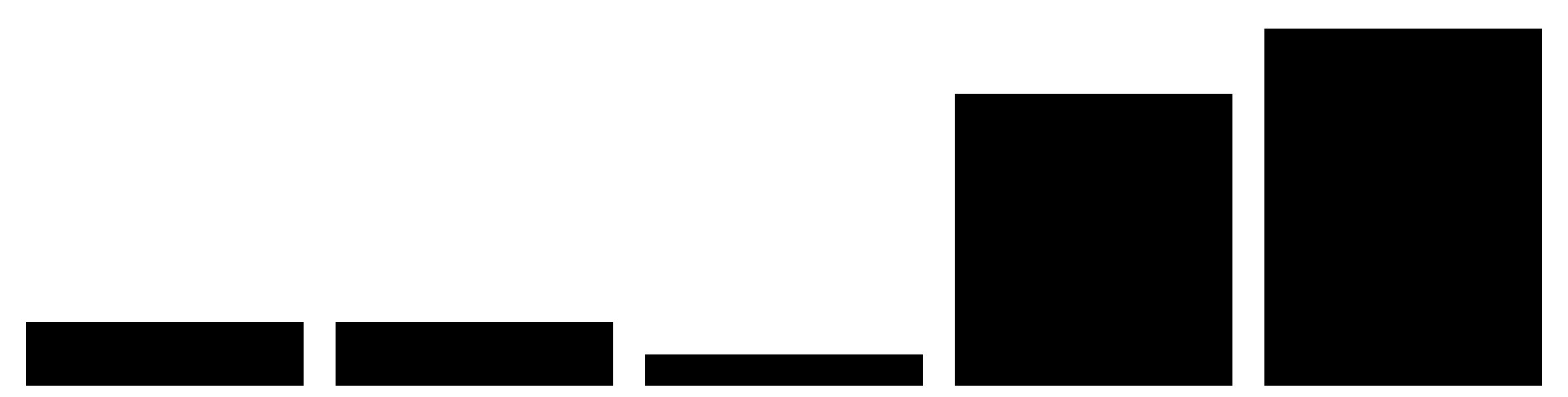}\\
\rowcollight Lack of Support & 3.96 & \edit{4} & 1.22 & \includegraphics[height=6pt]{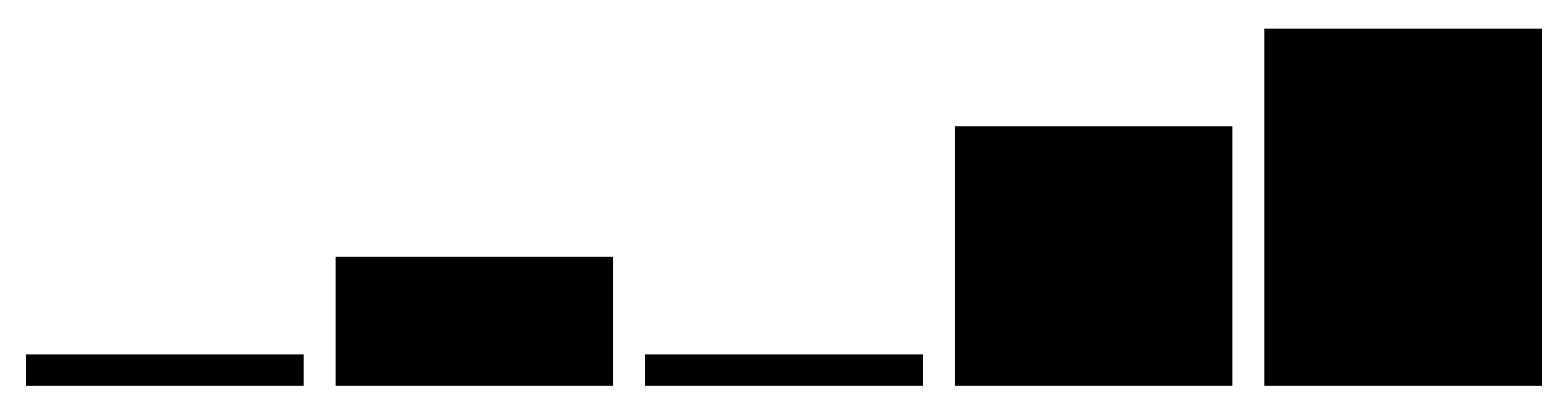}\\
Low Self-Efficacy & 3.16 & \edit{3} & 1.19 & \includegraphics[height=6pt]{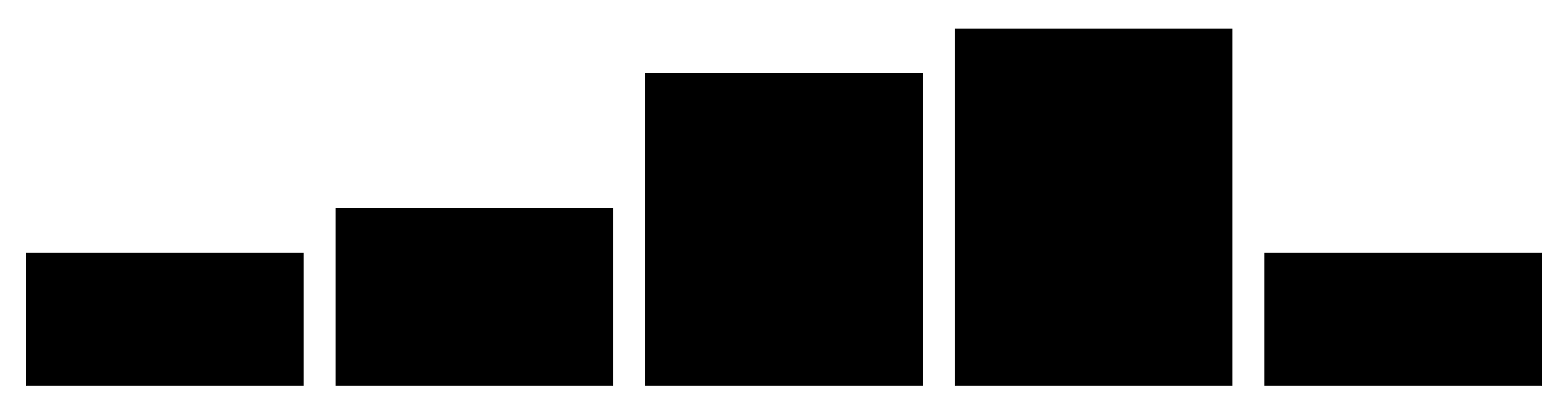}\\
\rowcollight Emotional Distress & 4.32 &\edit{5} & 0.97 & \includegraphics[height=6pt]{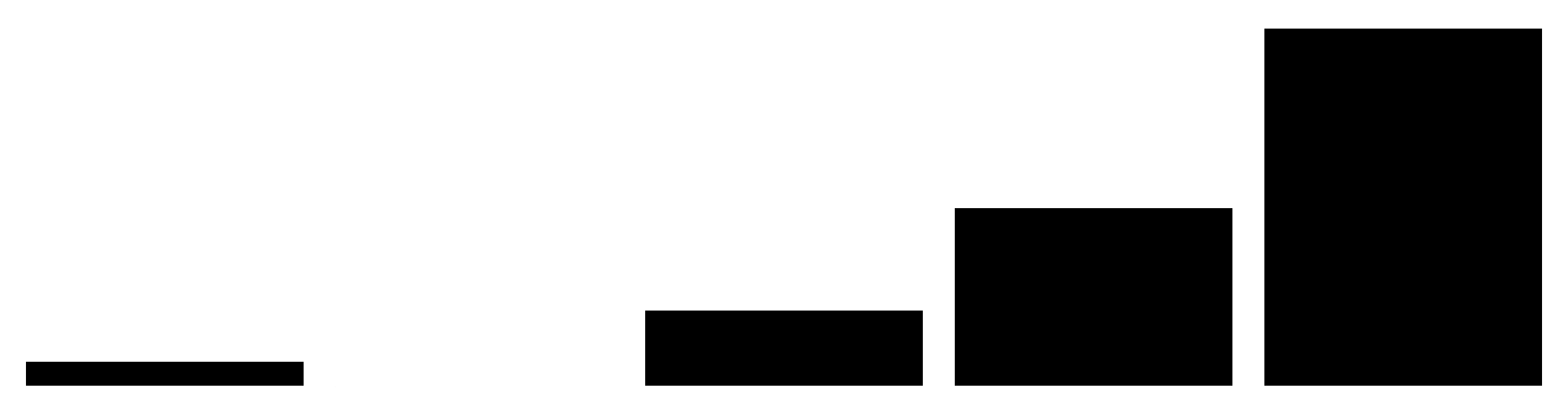}\\
Relationship Tensions & 3.56 & \edit{4} & 1.53 & \includegraphics[height=6pt]{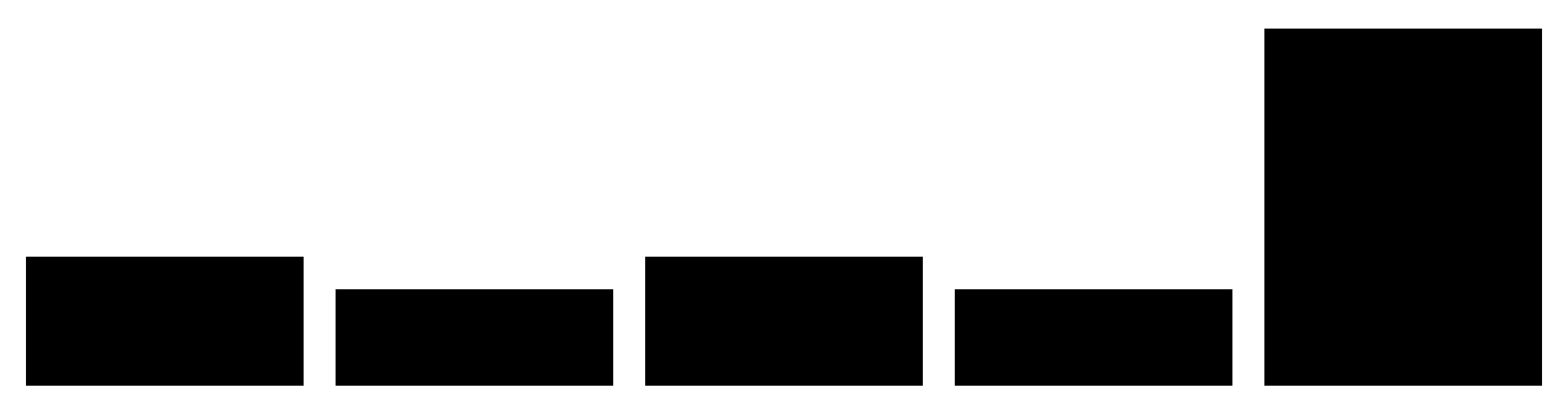}\\
\rowcollight Compassion Fatigue & 3.80 & \edit{4} & 1.30 & \includegraphics[height=6pt]{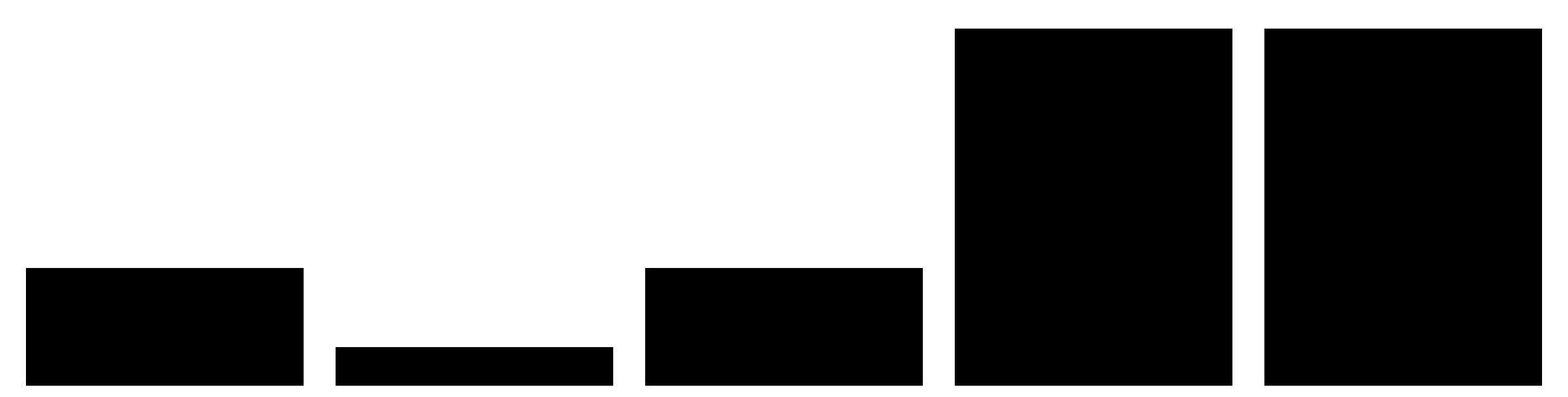}\\
Lack of Self-Care & 4.04 & \edit{5} &  1.30 & \includegraphics[height=6pt]{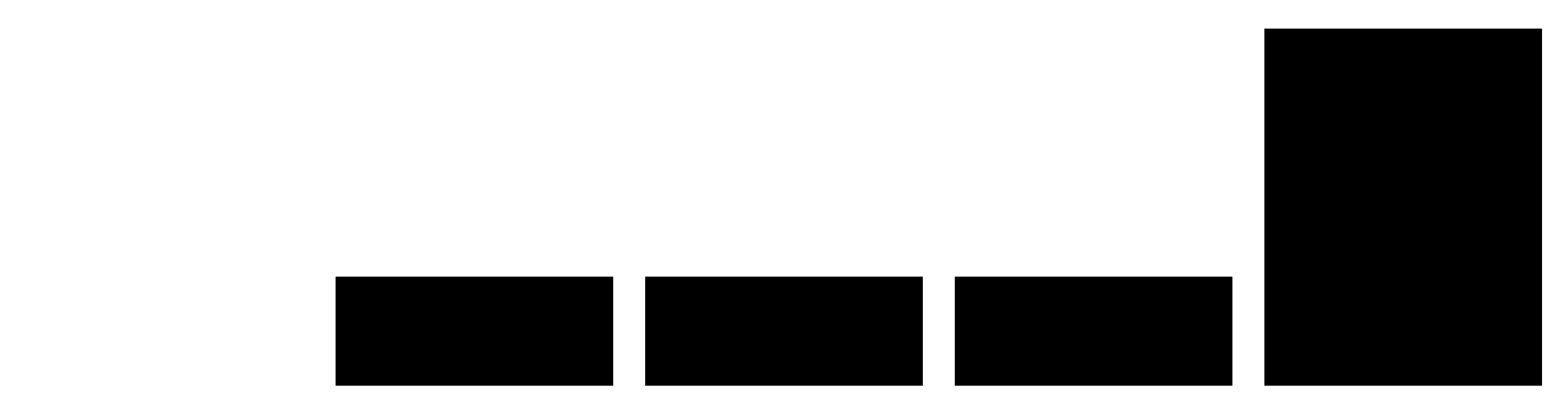}\\
\rowcollight Burnout & 4.24 & \edit{4} & 0.81 & \includegraphics[height=6pt]{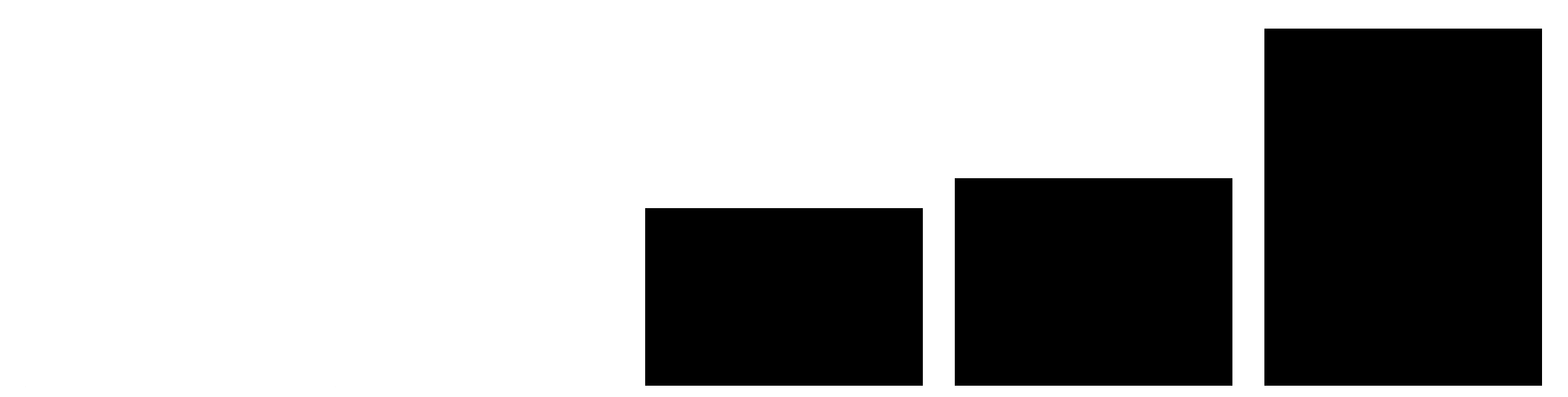}\\
\bottomrule
\end{tabular}
\end{table}

\subsection{Data Analysis}

% After the interviews, the research team used Otter.ai for the first five interviews and Microsoft Teams for the remaining interviews to automatically transcribe all audio recordings.
After the interviews, we used \textit{otter.ai} for automated transcription of the five interviews conducted on Zoom, and used the default transcription on Teams for the remaining interviews.
The recordings were anonymized by redacting any identifiable data such as personal names and locations.  
The dataset was then treated as a corpus for comprehensive analysis. 
% \edit{All five co-authors were involved in the open coding process. The co-first authors led the coding effort, completing the majority of coding on raw interview transcripts, while the senior co-authors (who have extensive experience in qualitative research) coded three raw transcripts each during two two-hour screen-sharing based co-working sessions to guide the learning process.}

% \edit{Our approach involved thematic analysis~\cite{braun2019reflecting} to identify patterns across participant experiences, with all participant quotes clearly labeled by ID throughout our findings. We selected our measures based on established frameworks in caregiving and mental health research, which we detail in the following sections.}
\edit{We analyzed our data comprehensively using reflexive thematic analysis~\cite{braun2019reflecting}. This analysis incorporated transcriptions from interview recordings, notes taken during interviews, and participants' responses to both the R-CWBS (\autoref{tab:participants}) and mental wellbeing concern prompts from the literature (\autoref{table:mentalwellbeing}). First, the co-first authors carefully reviewed each transcript against the original recordings to ensure accuracy and to capture nuanced expressions and emotional responses. Then, we organized our initial coding using Miro~\cite{miro2025} as a visual collaborative platform, which facilitated the identification of emerging patterns. Through this approach, we systematically developed our initial themes, which were then refined through team discussions and iterative analysis.}
%\edit{To analyze our data, we adopted inductive open-coding~\cite{} followed by thematic analysis~\cite{braun2019reflecting} to identify patterns across participant experiences.}
% \koustuv{@Melissa, this needs to be elaborated.} 
%We describe our approach below.

All co-authors participated in reviewing transcripts and engaged in an iterative open coding process, where codes were grouped, initial subthemes were identified, and these subthemes were refined into higher-level themes.
% \edit{To further elaborate our process, all five co-authors were involved in the open coding process. 
\edit{The co-first authors led the coding effort, completing the majority of coding on raw interview transcripts, while the senior co-authors (who have extensive experience in qualitative research) coded two raw transcripts each during two two-hours hybrid (in-person and screen-sharing-based) co-working sessions.}
To ensure coherence, themes were reviewed, merged, split, or discarded based on relevance to the RQs, resulting in the final thematic structure.
% in the analysis, we carefully reviewed and refined the themes. 
% This process involved merging certain themes into broader categories, separating other themes into distinct categories, and discarding themes that were not directly relevant to our core research questions. Ultimately, we determined the final themes corresponding to each research question. 
% While we allowed flexibility in our coding and theme development, our analysis was guided by prior research on social support in online dementia communities~\cite{johnson2022s}.
\edit{While we allowed flexibility in our coding and theme development, we employed thematic analysis~\cite{braun2019reflecting} as our primary methodological approach, guided by prior research on social support in online dementia communities~\cite{johnson2022s}}. 
A random subset of 10 transcripts was selected for an initial round of open coding. 
We developed 400 codes in the first iteration, which were then discussed by the entire team and distilled into 851 codes. 
The new set of codes were applied and iterated on the remaining transcripts. 
We discovered an additional 451 codes from the remaining transcripts.
The codes were then grouped into 12 higher-level and 63 lower-level themes, which aligned with our three RQs.

% \subsection{Ethics and Privacy}
\subsection{Privacy, Ethics, and Reflexivity}

Our study was approved by the Institutional Review Boards (IRBs) at the researchers' institutions.
Given the potentially sensitive nature of the study, we adopted several ethical and privacy considerations. 
To maintain confidentiality, each participant was assigned a unique participant ID to ensure anonymity. 
% to ensure adherence to ethical guidelines. 
Throughout the interviews, person-first language was used, referring to the loved one as a ``care recipient'', ``your father'', or ``your wife''  rather than an ``AD/ADRD patient'' to avoid any potential discomfort associated with clinical terms. 
During interviews, we monitored participants' emotional wellbeing, paying attention to both verbal and nonverbal cues and checking in directly to confirm their comfort and willingness to proceed. 
When participants became emotional while recalling memories with their loved ones, we would pause the interview and ask them to take a short break or have some water as needed.
Participants were also reminded of their right to discontinue the session at any time if they felt uncomfortable or no longer wished to continue.

Our research team comprises researchers holding diverse gender, racial, and cultural backgrounds, including people of color and immigrants, and hold interdisciplinary research expertise in the areas of HCI, CSCW, UbiComp, and Health Informatics. We have prior experience in working on the topics of mental health and wellbeing, AD/ADRD, and online social support. 
Multiple authors have served as caregivers for aging family members, although not specifically for AD/ADRD conditions.
While we have taken the utmost care to capture and faithfully synthesize participants' viewpoints, we acknowledge that our perspectives as researchers and, in some cases, as caregivers may influence our interpretations. 
We remain committed to conveying participants' experiences as authentically as possible and to highlighting the complexities of caregiving as voiced by those directly involved.

 %2 page
% \input{4RQ1} %2 page
\section{RQ1: Mental Wellbeing Needs and Concerns}\label{sec:rq1}

% \melissa{Finding part}
We identified a number of themes associated with the mental wellbeing needs and concerns of AD/ADRD caregivers. 
% Our analysis of the data led to several themes which despite being distinct were strongly related to each other. 
% In our pursuit to better communicate the relationships, w
We have adopted a categorization of these needs and concerns into two high-level themes of cause and effect---1) (Cause) Factors leading to mental wellbeing concerns, and 2) (Effects) impacts on the mental wellbeing of caregivers. 
The categorization in terms of cause and effect is meant to facilitate a better understanding of the relationship among the themes and not an interpretation of causality.
In this section, we first describe these themes (\autoref{sec:concerns}), followed by how the mental health concerns evolve over caregiving (\autoref{evolution}).

% \subsection{What are the primary mental wellbeing needs and concerns of AD/ADRD family caregiver?}
% \subsubsection{Factors Leading to Caregiving Challenge}

% \koustuv{2AC: consider merging 4.1 and 4.2, as the findings are well-supported by literature and not that novel.}

% \begin{figure}[t]

%     \centering
%         % \includegraphics[width=\columnwidth]{figures/Figure.pdf}
%         \includegraphics[width=\columnwidth]{figures/MentalHealthCauseEffects.pdf}
%     \caption{An overview of the sources and impacts of mental health concerns among caregivers as observed in our study.}
%     \label{fig:mentalhealth_concerns}
% \end{figure}

\subsection{Mental Wellbeing Concerns of Caregivers}\label{sec:concerns}
\edit{The family caregivers of individuals with AD/ADRD experience a complex, evolving set of mental wellbeing concerns. These challenges stem from multiple, interrelated sources and manifest in a range of social, physical, and psychological effects that shift over time. }
% \autoref{fig:mentalhealth_concerns} provides an overview of the sources and impacts of mental health concerns among the caregivers. 
% Importantly, these factors stem from their caregiving responsibilities.}

\subsubsection{Intersecting Sources of Caregivers' Distress} Participants consistently pointed to several overwhelming challenges, briefly described below:

\para{Financial Burden.} Several participants expressed financial burden as a major concern impacting their mental health. 
P5, P6, P7, and P18 noted how it creates uncertainty about care continuity. 
% Participants mentioned that there are not enough systems to provide financial support for caregiving; P6 mentioned that they had already started tapping into personal savings after exhausting all other financial resources.
With limited financial support systems, some, like P6, had to rely on personal savings. 
This instability added to caregivers' anxiety, e.g., P18 feared their funds might not last the care-recipient's lifetime.
% In fact, the stress of financial instability further contributed to anxiety and emotional distress among caregivers---P18 was worried that they may not have enough funds to last the lifetime of their care-recipient. 

\begin{quote}
\small
``I worry about having enough savings to last his life, he's 84. His mom lived to 94, so conceivably, he's healthy otherwise. 
He is physically fit, so he could conceivably live another ten years or more, and so I worry about his money lasting that long.'' ---P18
\end{quote}
% \ravi{see if we have a quote that explicitly says that finance/money is a contributer to mental health}

\para{Disrupted Social Life.}
Caregiving often disrupts caregivers' social life. 
Participants expressed losing touch with friends and having difficulty in participating in social activities. 
P1, P20, and P21 described struggles balancing caregiving with friendships; P1 explained ``I can't enjoy social interactions with my friend because I have a responsibility, and whenever I'm outside, I feel so anxious. I'm always thinking about the [care recipient].'' 
This sense of isolation intensified feelings of loneliness and emotional strain. P3 also noted a reluctance to open up for fear of judgment, choosing instead to share anonymously on social media.

% The demands of caregiving responsibility left them feeling isolated from social life, exacerbating feelings of loneliness and emotional strain. 
% Further, P3 expressed concern that they could not open up to others due to fear of being judged---therefore, they resorted to anonymous social media accounts:

\begin{quote}
\small
% I feel like I'm so disappointed because of the fact that 
``[..] I've sacrificed everything for [my mother]. I've stopped working, I lost my friends.
% , like, I feel like I've sacrificed a lot, a lot for her. So when she does that, 
I feel like giving up. I cry, I break down. I have no one to open up to, because I'm not comfortable sharing my problems with someone I know, because I feel like they may judge me [..]
% , or maybe when they come to our home, they'll start treating my mom in a bad way, depending on the Bad things I've been telling them about her, but sometimes I understand her. Maybe it's because of her condition, but you know, 
% [..] when I open up to someone, they may fail to understand her. 
% So that's the main challenge I face. I feel like I can't open up. 
I can't open up fully to someone I know. I'd rather go to social media and type what I'm going through using an anonymous account, and then maybe people will comment with legit and unbiased advice.''---P3
\end{quote}
% \ravi{missing PID for the quote}

\para{Time Constraints and Limited Personal/Self-care Time.}
Participants expressed continual struggle with time management, as caregiving demands often left little room for self-care.
 % due to caregiving demands, which often leaves little time for self-care.
Many felt overwhelmed, having to prioritize caregiving over their own wellbeing. 
P4 described caregiving as a \textit{highly time-consuming} job, explaining that factors such as commuting, unexpected incidents, and the care recipient's declining verbal abilities made it increasingly difficult to engage in other activities. P7 and P15 emphasized that caregiving often requires attention 24/7, leaving no time for breaks. Some, like P19, even changed jobs to better accommodate caregiving responsibilities:

\begin{quote}
\small
``As a school Superintendent, that was really, really stressful. 
I was responsible for taking care of thousands of people and I couldn't do that full time while also taking care of my mom full time.
So I decided to resign and taking a job with the university where I can work from home.'' ---P19
\end{quote}

\begin{quote}
\small
\edit{
``I was working in a store prior to that, so I had to resign to come in to take care of him. And then over the years, as I first started out like maybe two days a week, if I increased two days a week, now I'm a living caregiver.'' ---P22
}
\end{quote}

% \begin{quote}
% \small
% \edit{
% ``I've I stopped working, I lost my friends, like, I feel like I've sacrificed a lot, a lot for her.'' ---P3
% }
% \end{quote}

% Similar to the lack of social life, participants also voiced their concern about the lack of self-care or personal time. 
% A major concern expressed by caregivers was the significant loss of personal time. 
% P4, P7, and P15 reported that their caregiving responsibilities consumed nearly all of their time, leaving little to no opportunity for self-care. 
In addition, P10 expressed concern about being in a ``sandwich'' generation between caregiving and parenting, leaving no opportunity for self-care. 
Multiple participants expressed ``guilt'' about self-care, such as P23 felt guilty when they would take off time for themselves, ``I feel guilty trying to leave her alone and do other things.''
This lack of personal time resulted in increased stress and emotional fatigue, contributing to caregiver burnout among caregivers. 
P13 expressed that loss of travel has negatively impacted their mental wellbeing, with feelings of being stuck and having nothing to look forward to, and P24 expressed about being a \textit{prisoner in their home}.

\begin{quote}
    \small
    ``We're feeling like prisoners in our own home because we've discovered now that we can't leave [because of caregiving responsibilities].'' ---P24
\end{quote}

% \para{Worsening Care-recipients' Condition.}
% As AD/ADRD progresses, an individual's condition deteriorates over time. 
% Simultaneously, the caregivers experience heightened emotional strain and feelings of hopelessness. 
% As the care recipient loses memory and cognitive ability, they often forget their family member and cannot recognize the caregiver---which adds to the caregiver's mental health burden. 
% P11 described this situation as experiencing ``ambigious loss''---having their loved one physically present but no longer as they once were.

\para{Relationship Management and Tensions.}
\edit{Multiple caregivers expressed relationship tensions---whether from over-reliance or lack of support from family---as a key burden.}
Caregivers often faced imbalance when others did not share responsibilities, leading to stress and fractured relationships. P2 shared resentment toward an uninvolved sibling, while P18 worried about the toll on her marriage, saying her husband was supportive but felt ``thrown into this.'' 
% Many caregivers struggled to balance caregiving with maintaining family relationships.

% Relationship strain, including over-reliance by family members or conflicts arising from caregiving responsibilities, is a common source of burden. 
% This occurs especially when the caregivers rely on family members to share caregiving responsibilities but receive little to no reciprocation. 
% This can create an imbalance, placing disproportionate pressure on the caregiver, leading to increased stress and broken family relationships. 
% For instance, P2 mentioned, ``I had a sibling that didn't do anything to help, and so there was some resentment.'' 
% Again, P18 expressed continued worry about her marriage, noting that although her husband is supportive, she feels he has been ``thrown into this'' situation.
% Essentially, caregivers found a need to balance caregiving responsibilities with enough time for family members, such as:

\begin{quote}
\small
``[I'm worried] because apart from taking care of my mom, I have other family members.
% that I also have. 
I have a fiancee that I want to marry, so it has not been easy trying to balance taking care of my mom and taking care of my older and younger siblings, and my fiance. So I've been struggling to actually manage my relationships.'' ---P23
\end{quote}

% \begin{quote}
% ``I think so. I think so, especially, you know, I had a sibling that didn't do anything to help, and so there was some resentment. And then also with my dad, just he wouldn't really ask me if I had any plans or anything else that I needed to do. And so that was kind of a strain in our relationship, too, where he relied on me probably a little bit more than he should have. He wasn't willing to have other caregivers come into our house that we were all sharing. And so it was strictly me. And And so I think that it's kind of the number one detrimental thing to that mental health for me.
% ``P2
% \end{quote}

% \begin{quote}
% \small
% ``Engaged over, let's say problematically 24 hours because the job is quite demanding.
% Yeah, and can be so stressful most times.
% ``P7
% \end{quote}

\para{Overwhelming Caregiving Responsibilities.} 
% \koustuv{R3: Add more quotes and participant numbers here.}
Participants described feeling overwhelmed by the intensity and growing demands of caregiving, especially as the care recipient's condition worsened. Daily tasks included managing medications, finances, emotional support, and household responsibilities—often leaving caregivers exhausted. Many viewed caregiving not as a set of tasks but as a deep obligation (P1) or long-term commitment (P15).
\edit{For example, P25 shared that they would have to pick up each declining ability of the care-recipient:}

\begin{quote}
\small
\edit{
``So every time that he would lose an ability, I would pick it up. And so over time, that just gets more and more and more big because again, like for a three year old.'' ---P25 
}
\end{quote}

However, some like P18 share that responsibilities would not decrease despite using professional assistance or memory care:
% In addition, like P18, shared that their responsibilities increased after shifting their father to memory care, 
% as they now needed to constantly monitor the professionals' actions:

\begin{quote}
\small
``[It's] stressful having to basically supervise what these people are doing and they just don't have enough help in these facilities.'' ---P18 
\end{quote}

% \begin{quote}
% \small
% \edit{
% ''As their memory and physical ability declined, my role became more hands on and more demanding. Now I'm responsible for almost every aspect of their daily life, from personal care to managing their medications and to their safety as their disease advances.'' ---P21 
% }
% \end{quote}

% \begin{quote}
% \small
% \edit{
% ''So every time that he would lose an ability, I would pick it up. And so over time, that just gets more and more and more big because again, like for a three year old.'' ---P25 
% }
% \end{quote}

% \begin{quote}
% \small
% \edit{
% ''So as days go by, the condition worsens and I get more responsibilities, like before I used to like, just remind her of her routine. Now, sometimes I still I have to drive her, take her to shopping, take her to like, everywhere. Like, currently, it's like, I have my life and her life. I'm controlling my life and her life, because right now, she can't do anything by herself.'' ---P3
% }
% \end{quote}

Some participants also spoke about how the ``unpredictability'' with thier situations added to the emotional toll. They needed to constantly adjust their routines to the care recipient’s fluctuating condition, making it hard to plan and contributing to stress and anxiety:

\begin{quote}
\small
``One of the challenges I face is unpredictability which arises because the condition can actually change and change from day to day. So I'm always on high alert, and I'm always worried about their safety. And that can be quite exhausting.'' ---P20
\end{quote}

\para{Insufficient and Inefficient Support Systems.} Participants expressed the inadequacy of external support, whether from healthcare systems, community resources, government, or family. 
In particular, P2 emphasized that, while financial and healthcare benefits exist for patients, there is \textit{no} dedicated support system tailored to caregivers.
P4 noted that they use a plan that provides weekly support and resources, but had limited overall assistance and had the burden of paying for therapy out of pocket.
Further, governmental support avenues for care-recipients were deemed inefficient:

% eekly support and resources but notes limited overall assistance and the burden of paying for therapy out of pocket.(P4)

% In particular, P2 also highlighted that even though there are financial and healthcare benefits for patients, there is no stipulated support system tailored to caregivers.

% This lack of support leaves them feeling challenging. 
\begin{quote}
\small
``My dad is a veteran, and we've been trying to get him resources through the Veterans Administration, which has been horrible.  It is so hard to speak directly with someone, and resources have experienced roadblock after roadblock after Roadblock.'' ---P18
% trying to get him resources through the Veterans Administration. It has been horrible. It is so hard to find people to talk to and resources have experienced roadblock after roadblock after Roadblock.'' ---P18
\end{quote}

% , were worried about the care-recipient's condition if their own conditions deteriorated over time. 
% P4 was concerned about 
% Uncertainty regarding the future, both in terms of the care recipient's health and the caregiver's ability to cope, was a prevalent source of anxiety.

% \begin{quote}
% ``you know, self isolated, it, you know, like, social interaction, you know, hating sometimes, you know, like asking myself a lot of, a lot of question, you know, asking, you know, having, you know, being, being anxious, like, you know, I'll say like anxiety, Because I don't know how my future will be,
% ``P1
% \end{quote}

\subsubsection{Mental Health Impacts on Caregivers}
\label{effect}
The psychological toll of the above stressors manifested in multiple ways as listed below:

\para{Hopelessness about the Future.}
Many participants expressed anxiety and hopelessness about the future, both in terms of their ability to cope and the inevitable decline of the care recipient. 
Some, like P10, P11, and P20, worried about who would care for their loved one if their own health failed. P20 shared feeling overwhelmed after experiencing depression for the first time.
Others, like P4, described experiencing ``anticipatory grief''---a common theme in support groups---as they come to terms with the care recipient's ongoing decline and eventual passing.

% In fact, P10, P11, and P20 expressed concern about the condition of the care recipient should their own health decline over time. For example, when P20 experienced depression for the first time, they were scared about the future.
% In addition, P4 expressed that they talk about ``anticipatory grief'' in support groups---they suffer in silence that the care-recipient's condition will continue deteriorating and they will eventually pass away:

\begin{quote}
\small
The reality is that this is the only disease without a cure, and this disease is fatal, so we're dealing a lot with people with anticipatory grief, which is what we talk about a lot. [..] And many times, people are not going to support groups until they're like, literally in tears or at wit's end. I see anger from men and frustration, and I see tears and physical demise from women. ---P4
\end{quote}

\para{Fatigue, Strain, and Burnout.} 
\edit{Participants described fatigue and physical strain from the constant demands of caregiving. P1 noted ``the emotional toll of seeing a loved one suffer.''
and P17 highlighted both emotional and physical exhaustion. 
Some participants also reported the constant demand of caregiving responsibilities also led to a lack of sleep and chronic sleep deprivations. 
Many reported experiencing burnout, often realizing only in hindsight how deeply caregiving had affected them.}

% Participants reported experiencing overwhelming fatigue and physical strain due to the round-the-clock demands of caregiving and the care recipient's condition. 
% P1 expressed, ``the emotional toll of seeing a loved one suffer,'' and P17 mentioned, ``caregivers face emotional strain from care recipient's decline, physical exhaustion.''
% Many caregivers experienced burnout---a state of complete physical and mental exhaustion.
% Participants also recalled, in hindsight, that they were often unaware of the emotional toll caregiving was taking on them:

\begin{quote}
\small
``I feel increasing [caregiving] demands, my physical and emotional demand. This affected my mental health. I became more concerned about burnout and how to maintain my wellbeing.''---P21
\end{quote}

\para{Emotional Upheavals and Compassion Fatigue.}
The overwhelmingness of caregiving also leads to frequent emotional upheavals. 
These emotions are often triggered by the care recipients' health decline, unpredictable behaviors, and strain of balancing caregiving and other responsibilities:
% or the strain of balancing caregiving with other responsibilities, such as:
% P22 described caregiving as a ``rollercoaster of emotions.'' 
% P22: "Caregiving like a rollercoaster of emotion for me."

\begin{quote}
\small
``It hasn't been easy for me because I get very anxious [..] prepare for the worst [..] there are some days that I really, really hope for the best. So it has been like a rollercoaster of emotions.''---P22
\end{quote}

Prior work notes compassion fatigue among caregivers of chronic conditions---compassion fatigue occurs when the caregiver's ability to empathize with the care recipient is reduced as a result of repeated exposure to their suffering~\cite{day2014compassion}.
Similarly, compassion fatigue emerged as a major theme in our caregivers' experiences. 
% Compassion fatigue among caregivers reflects the emotional exhaustion that builds from long-term exposure to care recipient's suffering and constant caregiving responsibilities. 
% For instance, P6 explained the sense of hopelessness that often accompanies compassion fatigue.
% \begin{quote}
% \small
% \melissa{not sure if this is a good quote for compassion fatigue}
% ``I  feel that way because especially when you know that hopelessness comes in when you feel you've been you've been you've been so compassionate about the person and it looks like nothing is happening.''---P6
% \end{quote}
For instance, P10 explained they were often drained from caregiving, and would even lose their temper when the care recipient would turn aggressive:

\begin{quote}
\small
``I think I did a good job of reminding myself that when my mom was like, really bad, that it's not her, it's the disease, and it's her brain not working right. It's not her choice. But sometimes she was just [****]! And so I lost my temper.''---P10
\end{quote}

% This responses illustrates the intense emotional toll that compassion fatigue can have on caregivers, complicating their mental wellbeing and ability to provide empathic and patient care. 

\para{Self-Reflective Positive Impacts.}
In addition to several negative psychological impacts of caregiving, some participants also reflected on certain positives they drew out of caregiving demands. 
For example, P1 observed personal growth, noting they had become more mature over time. 
P11 and P18 shared that \edit{navigating daily caregiving challenges} helped them develop self-efficacy, as they recognized their growing competence in caring for others. 
Similarly, P6 reflected on becoming more compassionate, detail-oriented, and gaining a deeper understanding of others:
% personal growth and had a better understanding of others:
% \begin{quote}
% \small
% ``I think [caregiving] has both positive and negative impact. [for positive], you have to be there full time, you have to always be there and then helping out and all of those stuff and you happen to see this person is really not not recovering at the peace which you would expect the person to [..] Because you're wanting the best for this person and also it's it's it's it's also in mental health positively by helping you to know how people reason such that you I I find myself not having issues with people because it caregiving activities has helped me.
% ''---P6
% \end{quote}
\begin{quote}
\small
``One thing I've learned as a caregiver that impacted my mental health a lot is that now I seek first to understand before being understood. I don't wanna impose my decisions on people.
I want to know why they do the things they do, why they say the things they say before I respond.''---P6
\end{quote}

\edit{The above reflections suggest that, for some caregivers, the intense demands of caregiving also fostered a sense of purpose, personal growth, and emotional depth---underscoring the complex, dual nature of caregiving journey.}

\edit{Overall, the caregiving-related concerns identified in prior research served as useful starting points for our conversations with participants (\autoref{table:mentalwellbeing}). These prompts helped guide participants to recall experiences related to emotional distress, burnout, and lack of support---all of which strongly resonated in their narratives. Many described feeling overwhelmed and neglecting their own wellbeing due to caregiving demands.
Our findings also surfaced organically emerging themes, such as financial stress, loss of personal and social life, strained relationships, guilt around self-care, and limited family support---issues not always explicitly framed as mental health concerns in prior work. 
Notably, participants also shared self-reflective positives, including personal growth and emotional resilience, expanding the discourse beyond problem-focused narratives.}
\subsection{The Evolution of Mental Health of Caregivers}
% \edit{Highlight the evolution of caregiving needs}

\label{evolution}

\begin{figure}[t]

% \melissa{Can we delete the last row "Practices & Coping Strategies" and create a separate table "Mapping of Mental Wellbeing Practices to Stages of Caregiver Mental Health Evolution"}
    \centering
        \includegraphics[width=\columnwidth]{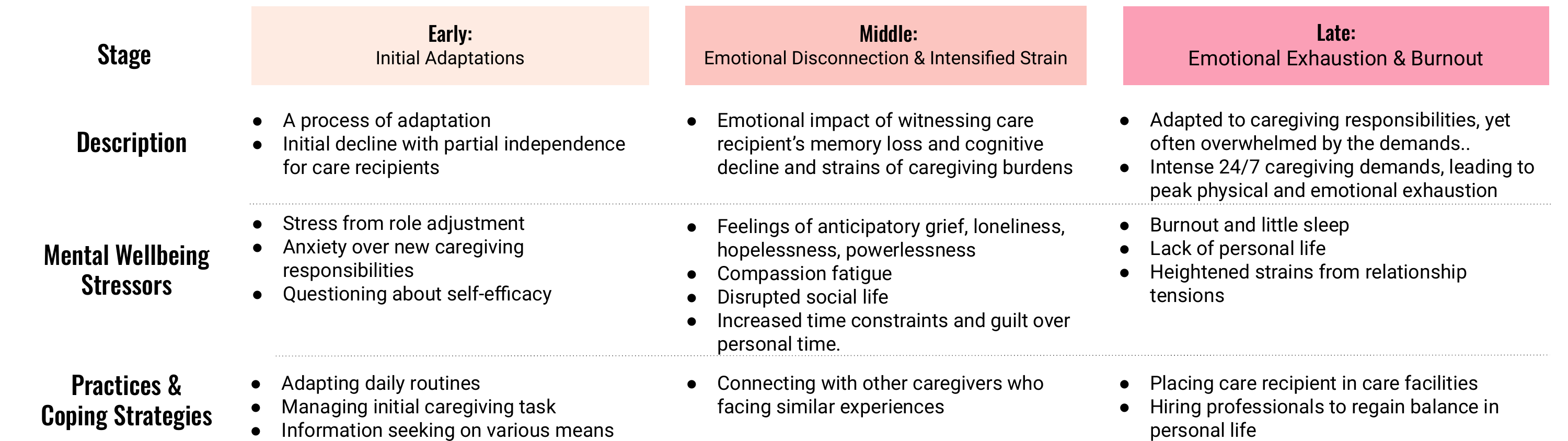}
    \caption{An overview of the evolution of mental health concerns experienced throughout AD/ADRD caregiving.}
    \label{fig:evolution}
\end{figure}
% Through the interviews we conducted, w
Given that the manifestation of AD/ADRD can vary from individual to individual, the caregiving experiences also differ. 
However, based on our interviews, we discovered some high-level patterns that  we have distilled into three stages for better understanding some key challenges and exploring the potential of technology to address them.
%Based on our interview, 
We characterize the evolution of caregivers' mental health into---
1) Initial Adaptation, 2) Emotional Disconnection and Intensified Strain, and 3) Emotional Exhaustion and Burnout.
\autoref{fig:evolution} provides an overview of the different stages and their associated key stressors and coping strategies.
Stressors and coping strategies highlighted in the figure are assigned to the stage where they most commonly appear.  

\para{Early Stage: Initial Adaptations, \edit{Shock, and Uncertainty about Future}.} 
The early stage of caregiving is characterized by a process of \textit{adaptation} \edit{as individuals adjust to the new and often overwhelming responsibilities for caring for someone with AD/ADRD.}
\edit{Caregivers reported often underestimating the emotional impact of caregiving.}

\begin{quote}
\small
\edit{``When I first became a caregiver, I didn't realize how much time and energy would be consumed by caregiving. I was in denial about how much this would impact on me emotionally.''---P21}
\end{quote}

This phase typically coincides with the care recipient's initial decline in health, where care-recipients can still manage many activities independently.
% However, there is a transition in the caregivers' lives, including a disruption of their established routines.
% Most caregivers who begin their caregiving journey at this point often face a significant change in their daily lives. 
% During this stage, they experience stress and anxiety as they need to transition from their (previously) usual routines to becoming caregivers. 
\edit{Following the diagnosis of the care-recipients, caregivers are likely to experience intense stress, panic, and anxiety, driven by the uncertainty about the future and the evolving demands of the role.
Some caregivers reported the ``role shock''---feeling unprepared and emotionally unsteady as they stepped into caregiving.
During this time, caregivers may struggle with self-efficacy, doubting their ability to provide effective care, which can further contribute to feelings of depression and anxiety:}
% feel insufficiently prepared, feeling depressed and anxious:}

\begin{quote}
\small
\edit{``In the beginning, I was constantly feeling depressed while seeing them.[..] That first year was especially difficult. I felt overwhelmed and nearly always in a state of depression.''---P15}
\end{quote}

\edit{The initial shock and the addition of new responsibilities was often overwhelming, leaving no time for the caregiver to reflect on their own health and needs, as expressed by P2:}

\begin{quote}
\small
\edit{``I didn't really know how to manage patients with Alzheimer's. I didn't take breaks [..] I think when you're in the middle of it, you just adapt and don't really think about it that much.''---P2}
\end{quote}

% \begin{quote}
% \small
% \edit{``I never knew what I was going to actually meet when trying to be a caregiver...It was shocking to me to see that it wasn't a short term, it was a long time stuff.''---P16}
% \end{quote}

% In addition, caregivers also experience challenges to their self-efficacy, doubting their abilities to provide effective care.
% This phase typically coincides with the care recipient's initial decline in health, where care recipients can still manage many activities independently.

% \ravi{Do we have data about stressors specific to early stage? e.g., needing to take time off from work}
% \ravi{Do we have any quote that says that caregivers do not need much help in the early stages?}

% where caregiving responsibilities are relatively minimal, and the care recipient can still manage some activities independently. 
% \begin{quote}
% \small
% ``Well, early on, she was still driving. She was still going to grocery store. She would. You know pretty, fairly independent, although she was forgetting things, forgetting appointments, she was still functional and society.''---P13
% \end{quote}

\para{Middle Stage: Emotional Disconnection and Intensified Strain.}
%The first stage of caregiving is characterized by a process of adaptation as caregivers adjust to their new responsibilities. During this phase, the emotional strain on caregivers intensifies, with many citing this as the most difficult stage of their journey. One of the most distressing experiences for caregiver is when their loved ones can no longer recognize them due to memory loss. This emotional disconnection was deeply painful for caregiver(P12). In some cases, caregivers had to cope with care recipient's unpredictable behavior, such as aggressive outbursts(P20), knowing that the care recipient would not remember these incidents the next day(P24). The anxiety was compounded when medical evaluations confirmed the worsening condition of the care recipient, making caregivers feel even more powerless (P22).
\edit{Even after the initial shock of the diagnosis and responsibilities subsides, the caregivers typically do not get a long break. 
As the care-recipient's condition worsens, caregivers face growing emotional burdens, feelings of isolation, and need for support and recognition.}
% deals with increasing emotional burden, feelings of isolation, and recognition for support.}
% The progression of caregivers' mental health challenges often stems from the emotional impact of witnessing their loved one's memory loss and cognitive decline. 
\edit{In the mid-to-late stages of caregiving, recurring episodes of the care-recipient's decline or emotional instability can lead caregivers to experience intense moments of emotional breakdown (P3, P21).}
They can experience deep emotional loss, as they feel they are losing the person they once knew, and can often feel grief, loneliness, and hopelessness. 

\begin{quote}
\small
\edit{``As I formed them deeper emotional bonds the mental toll became more evident.''---P17}
\end{quote}

The participants described the feelings of ``anticipatory grief'' and ``ambiguous loss'' as they came to realize that the care recipient's condition would not improve and that the person they once knew was no longer the same.
Multiple participants expressed feeling powerless that despite their efforts, they are unable to protect their loved one from the disease's progression, which further deteriorates their mental wellbeing.
% The reason for the evolution of mental health is largely due to the emotional devastation of realizing that PwD's memory loss. This represent not just a cognitive decline but and emotional loss, where caregivers feel as though they are losing the person they once know. The disconnect between the caregiver’s deep emotional investment and the care recipient’s fading recognition creates profound feelings of grief, loneliness, and helplessness. Caregivers may feel that despite their efforts, they are unable to protect their loved one from the disease's progression, which further erodes their mental wellbeing.
At this stage, caregivers need emotional support. 
Connecting with others going through similar experiences can provide comfort and validation: 
% Participants such as P11, P21, and P24, mentioned the online platforms, such as Reddit. They can seek support groups where they can share their challenges and emotions. In these online communities, caregivers can engage with people who have faced similar struggles, offering them a safe space to vent their frustrations, ask for advice, and receive emotional reinforcement. The ability to connect with others who truly understand the complexity and pain of their caregiving journey helps alleviate feelings of isolation, providing a sense of community and shared understanding during this emotionally taxing phase. 

% \begin{quote}
% \small
% ``So my mental health has fluctuated over time.
% At first, I was in denial about how much this would impact me emotionally. Over time, I have had to seek out more mental health resources to cope with stress and the emotional and emotional strain.''---P21
% \end{quote}

\begin{quote}
\small
\edit{``I started to like experience, feelings of burnout and stress. I found out that I needed more emotional supports and not just from my colleagues, but also from therapy or peer groups.''---P17}
\end{quote}

% This period also coincides with increased disruption in caregivers' social and personal lives, as well as feelings of \edit{guilt} over taking personal time off from caregiving responsibilities. 
\edit{This is also the time where caregivers have been in their new role for a while and their new responsibilities start to put a strain on their social life. 
Social stigma around the condition also made it challenging for some caregivers to maintain a healthy social life.}

% \begin{quote}
% \small
% \edit{``It felt isolating because none of my friends really understood what it was like to take care of someone.[..] It's very, very easy to lose touch with friends and social activity because so much of my focus is on providing care.''---P21}
% \end{quote}

\begin{quote}
\small
\edit{``Now, I have no one to open up to. I'm not comfortable sharing my problems to someone I know. I feel like they may judge me---or treat my mom differently when they visit our home.''---P3}
\end{quote}

\edit{Some participants also expressed frustration about the lack of dedicated professional assistance for dementia caregivers, in terms of the lack of relatable experience or empathy, e.g.,:}

\begin{quote}
\small
\edit{``I think the person that I was talking to just didn't quite have the much experience in helping people … it's kind of hard to relate to that person.''---P2}
\end{quote}

% \edit{In addition, participants also face the fear of judgment from social and familiar circles, and may start self-isolation and withdrawal to avoid such interactions, e.g,:}

% \begin{quote}
% \small
% \edit{``Now, I have no one to open up to. I'm not comfortable sharing my problems to someone I know. I feel like they may judge me---or treat my mom differently when they visit our home.''---P3}
% \end{quote}

\edit{Between the middle and late stages, caregivers often realize that they are in a marathon and not a sprint. This is when they start reaching out for support through online or in-person communities.}

% \begin{quote}
% \small
% \edit{``I now make a point to communicate with family or friends when I need a break, whether that having someone step in to care for a few hours or scheduling a respite care.''---P21}
% \end{quote}

\begin{quote}
\small
\edit{``I use technologies on social online platforms like Reddit and also various apps to manage my mental health because they offer support and resources that fit into my busy schedule.''---P17}
\end{quote}

\para{Late Stage: Emotional Exhaustion and Burnout.}
Over the significant progression of time and \edit{the decline of} care recipient's condition, caregivers adapt to their caregiving duties almost like a full-time job. 
% \koustuv{@melissa, could you revise this part: please add stuff about what participants said aboud burnout. You can also say that some caregivers put their PwD facilities. Some caregivers prioritize about mental wellbeing whereas some stop thinking about themselves any longer. You can talk more about anticipatory grief here. }
% In the last stage of caregiving, caregivers often find themselves responsible for their loved ones 24/7, leaving no time for their own needs. Caregivers report being overwhelmed by the increase responsibilities, such as managing their bill, estate, and doctor appointments (P20), while others describe the strain of fully taking care of all aspects of the care recipient's life (P1). 
% The reason for mental health decline in this stage is that caregiving task occupy most of their time, with no time left for the caregiver to rest or focus on their own wellbeing. Caregivers feel as though they have no escape from their responsibilities, which drastically impacts their mental health.
% In the late stage of caregiving, m
However, by this point, they are also overwhelmed by providing 24/7 care, leaving little to no time for self-care.
% This leaves them with no personal time to address their own needs. 
This period marks peak physical and emotional exhaustion, with burnout becoming a frequent concern.
Participants mentioned severe sleep deprivation, as P11 described they would almost ``sleep with one eye open.''
P10 described the demands of managing every aspect of the care recipient's life, including handling bills and estate matters, coordinating constant medical appointments, and providing hands-on care and support.

This stage also coincides with heightened strains from relationship tensions, such as unmet expectations about sharing caregiving responsibilities among family members or concerns about the impact of caregiving demands on other relationships (e.g.,spouse and children), such as:

\begin{quote}
\small
\edit{``I had a sibling that didn't do anything to help which caused a lot of resentment. My dad never really asked if I had other commitments—he wasn’t open to bringing in outside help, so the responsibility fell entirely on me. That lack of family support was probably the biggest factor that was detrimental to my mental health at that time.''---P2}
\end{quote}

% \begin{quote}
% \small
% ``I do all his bills for him now [..] I have power of attorney, such as managing his estate, the assisted living place [..] but everything else related to his. I take him to doctor's appointments that for any of the doctors that he sees that don't come to the assisted living facility.''---P10
% \end{quote}
The all-encompassing nature of caregiving at this stage means caregivers can reach a critical point where they make difficult decisions regarding the care recipient, such as placing them in memory care facilities or hiring professionals to regain some balance in personal lives:
% . Some decide to place Pwd in a care facility to manage their own mental health and regain balance in their personal lives. 
\begin{quote}
\small
``And then we made the decision to put her into memory care. And so memory care has eased a lot of the burden on me and my family [..] 
% then when she went to memory care, it really relieved a lot of that pressure [..] 
now that she's in memory care now it's just about more of playing an advocacy role.
''---P19
\end{quote}

% For these caregivers, prioritizing their mental health becomes essential to sustain their wellbeing and recover from the long term stress. 

\para{\edit{Experience/Journey of Serial Caregivers.}} This three-stage process primarily reflects the typical experience of first-time caregivers. For \textit{serial caregivers}---those who have cared for multiple family members with AD/ADRD, like P2, P16, P17, P21, and P22---the experience evolves across iterations. Technological advancements also shape these experiences; for instance, P2 noted that smart home technologies would have made a significant difference during their caregiving two decades ago.

% A key consideration to note about this three-stage process is that it primarily represents the typical experience of a first-time caregiver.
% For people who are \textit{serial caregivers} (i.e., they have cared for multiple family members with AD/ADRD), such as P2, P16, P17, P21, and P22 in our study, the experience changes over iterations. 
% That said, the rate at which technology is evolving also affects these experiences. 
% For instance, P2 mentioned that having smart home technologies earlier, would have had a significant impact when they were caregiving two decades ago.

\begin{quote}
\small
``I think back, if I had had the technology that I have now, things would have been a lot easier. I have, pretty much automated my house with cameras and front door locking and unlocking abilities. So I think now it would be a million times easier using technology to manage dealing with my mother, especially.
''---P2
\end{quote}

\edit{Similarly, P17 described how their caregiving approach shifted over time from mastering daily routines to becoming more emotionally attuned and better at setting boundaries. They overcame feelings of guilt about self-care over time:}

\begin{quote}
\small
\edit{
``Early on when I first started, I used to feel guilty about taking time for myself, but now I understand that self-care is essential for long-term experiences [..] my experience has shifted from task-focused to a more balanced approach.''---P17}
\end{quote}

\edit{
These reflections reveal that repeated caregiving experiences can help caregivers to adaptively grow, in terms of developing confidence, refining coping strategies, and becoming more intentional about balancing caregiving and self-care. Early caregiving experiences were often marked by stress and self-doubt, and they were focused on getting through daily responsibilities. However, in later iterations, caregivers developed a more holistic perspective, applying emotional insight, prior knowledge, and strategic use of resources to improve both care recipients and their own wellbeing. A key characteristic of a \textit{matured} caregiver was their development of \textit{resilience}: }

\begin{quote}
\small
\edit{``On the positive side, I've become more obviously more resilient, some better at managing stress [..] I've also learned coping mechanism like mindfulness and certain emotional boundaries that have also helped me stay grounded.''---P17}
\end{quote}

\begin{quote}
\small
\edit{``But now it has become much more demanding as they have lost more independence over time. I've learned to adapt and I've learned to become more patient.''---P21}
\end{quote}

\edit{
In this way, serial caregiver not only refined their practical skills but also redefined their caregiving identity, from overwhelmed responders to more empowered and balanced care partners. }

% \koustuv{This is for someone who is going thru the process the first time, but there were people who have had this for multiple times, but their experiences were different. But they have also gone through this in different technological eras, e.g., one participant mentioned that if some technologies existed back in the day, then their caregiving would have been different. How technologies play different role at different stages of caregiving} %2 page
\section{RQ2: Practices to Address Mental Wellbeing Needs}\label{sec:practices}

\edit{As caregiving responsibilities and mental health challenges evolve, caregivers adopt a range of wellbeing practices. These practices are often ad-hoc, shaped by immediate needs and lived experiences, but can also develop into consistent routines over time. In addition, they are not static; rather, they shift in response to changing emotional, physical, and relational demands of caregiving. \autoref{table:mental-health-practices-evolution} provides an overview of mapping mental wellbeing practices with the different stages of caregiving. 
In the early stage, caregivers navigate the initial shock and uncertainty of their new responsibilities, relying on institutional resources, physical activities, and online resources to gather information and regain control. 
In the middle stage, as emotional strain deepens, they expand coping strategies---seeking professional help, joining support groups, pursuing hobbies, and leaning on family for emotional and logistical support. In the late stage, marked by burnout, caregivers emphasize structured mental health interventions, firmer boundaries, and support from close networks.
% As emotional disconnection and intensified strain emerged in the middle stage, caregivers expand their coping strategies by seeking professional support, engaging in personal hobbies, participating in online support communities, setting firmer boundaries in daily life, and increasingly relying on family members for emotional and logistical assistance. 
% In the late stage, marked by exhaustion and burnout, caregivers emphasize the importance of structured mental health interventions, stronger boundary-setting, and leveraging support from family and close friends. 
This temporal framing underscores the need for flexible, stage-sensitive support systems.}
% highlights the adaptive nature of wellbeing practices and underscores the need for flexible, stage-sensitive support systems that align with caregivers’ evolving emotional and practical challenges.} 
Across all stages, we found two major themes of practices---1) \textit{seeking external support} and 2) \textit{adopting self-care practices}. 

\begin{table}[t!]
\centering
\sffamily
\footnotesize
\caption{\edit{Understanding the Evolution of Caregiver Mental Health through Wellbeing Practices
% Mental Wellbeing Practices Mapped to Each Stage of Caregiver Mental Health Evolution
}} 
% \koustuv{Editing this. Could be the final version. @melissa, could you add the example quotes here?}}
\label{table:mental-health-practices-evolution}
\edit{
\begin{minipage}{\columnwidth}
\begin{tabular}{p{0.4\columnwidth}p{0.59\columnwidth}}
\textbf{\edit{Mental Wellbeing Practices}} & \textbf{Example Participant Quote Snippet}\\
\toprule
\rowcolmedium \multicolumn{2}{c}{\textbf{Early Stage: Initial Adaptations, Shock, and Uncertainty about Future}}\\
\textbf{Institutional Support}: Looking up available resources and assistance programs, such as exploring the local Alzheimer Association support groups. & ``I logged on to a local we're here in the suburbs of North Atlanta, and I've logged on to a caregiver type Alzheimer's Association.''---P13\\
\hdashline
\rowcollight \textbf{Physical Activities}: Simple walks with the care recipient, basic stretching, deep breathing, and exercise. & ``We have a park nearby where we walk the dogs. We do a lot of swimming and and that sort of thing, some yoga and stretches.''---P2\\
\hdashline
\textbf{Online Resources:} Beginning the use of wellbeing and meditation apps, and participating in online communities to learn more about similar circumstances from community members &  
% Participating in online forums and using wellbeing/meditation apps (P17).
``The [meditation] app assists me and tells me to take some deep breaths at times. So once in a while too, I also go online to edit support groups, like I mentioned earlier, so I get advice from other caregivers on managing daily challenges.''---P20\\
% ''And apps like headspace or cam, they offer me meditation, breathing exercises, mindfulness practice. They provide different programs focus on reducing my stress on my side''(P21)\\
\rowcolmedium \multicolumn{2}{c}{\textbf{Middle Stage: Emotional Disconnection and Intensified Strain}}\\
\textbf{Professional Support}: Exploring therapy options, and setting up initial consultations to cope with stress and emotional strains. & 
``So I did do some therapy, off and on, for various reasons, mostly because I had three adults I was caring for at the same time.''---P5
\\
\hdashline
\rowcollight \textbf{Hobbies}: Reading novels, playing video games. & 
``It's helpful because of the fact that your mind is, my mind is always, always occupied. So if I'm free, if I'm not doing anything, I'm playing video games.''---P3 \\
\hdashline 
\textbf{Social Support from Online Communities}: Regular participation in online communities, engaging with peers. &
``So on Reddit, there are a lot of people that caregivers, such as I. So we actually have, like, we have a group where we actually discuss so on that group, everybody comes with their own challenges and how they face them. ''---P20
\\
\hdashline
\rowcollight \textbf{Setting Boundaries in Life}: Learning to decline additional responsibilities, creating physical separation, and prioritizing self-care activities. & 
``I've also become better at setting my own boundaries and prioritizing my own well. Being early on, when I first started, I used to feel guilty about taking time for myself, but now I understand that some sale cases are essential for long term caregiving.''---P17
\\
\hdashline
\textbf{Family Support}: Relying on family support to continue caregiving, and deal with difficult circumstances.  & 
``So those days I needed to, like, take care of myself, go out with friends and get things for myself, and then I also have my family come around to help, especially when it's proven difficult''---P22
\\
\hdashline
\rowcolmedium \multicolumn{2}{c}{\textbf{Late Stage: Emotional Exhaustion and Burnout}}\\
\textbf{Professional Support}: Regular therapy sessions, counseling, and professional mental health interventions & ``I needed ongoing therapy to help manage the caregiver stress at this stage.''---P19 \\
\hdashline
\rowcollight \textbf{Setting Boundaries}: Major boundary decisions (facility placement), clear separation of life aspects. & ``Deciding to send the care recipient to memory care, more time for the caregiver.''---P19 \\
\hdashline
\textbf{Family support}: Support from family and close friends.  & 
''But now I've come to realize that I can't do it all alone, and now I now I make a point to communicate with family or friends when I need a break, whether that having someone step in to care for a few hours ''---P21
\\
\hdashline
\rowcollight \textbf{Planning}: Comprehensive care planning, professional care management. & ``Stay organized, prioritize tasks, take short breaks'' ---P17\\
\bottomrule
\end{tabular}
\end{minipage}
}
\end{table}
%TC:endignore

% Based on the interview study we conducted, we identified two practices caregivers address mental wellebing needs. 

% \koustuv{2AC: Findings either obvious or lack originality. And then see how these practices could map to the evolution of caregiving needs.}

\subsection{Caregivers Seeking External Support}
% To cope with caregiving demands and manage their mental health, c
Caregivers emphasized the importance of seeking various forms of external support, including assistance from family members, professionals, and institutions such as employers, community resources, and governmental programs. 

% including assistance from professionals, institutional 
% highlighted the need to seek several forms of external support, including family, community, government, and professionals. 

% The participants highlighted the need to seek external support to overcome the caregiving demands as well as to manage their  

% Caregivers mentioned external sources of support to manage the emotional and physical toll that caregiving demands, such as Society and Government Support, Family Support and Professional Support. 

\para{Family Support.}  
Given that our study centers around family caregivers, several participants emphasized the importance of family support in managing their wellbeing.
% essential for caregivers, especially in our study, where most participants are family members caring for their own relatives. 
%Family support ensures caregivers do not feel solely responsible, fostering a collaborative approach that is essential for sustaining their wellbeing.
Support from family not only eases the caregiving burden but also offers vital emotional reinforcement. 
%P4 and P10 noted the important role of their spouses in managing household responsibilities and looking after their children while they are busy with caregiving for their respective elderly care recipients.
%They highlighted that their spouse would often remind them of self-care, which could otherwise be easily ignored. 
% In fact, P10's husband would also frequently check the data of the (Oura) smart ring to ensure that P10 is getting enough sleep. 
% In addition, 
P24 shared that their brother and wife occasionally step in to share caregiving duties, providing them with much-needed breaks.

% Family encouragement plays a vital role for caregivers, particularly in our study, where most participants are family caregivers caring for their own relatives. Support from family members not only lightens the caregiving load but also provides emotional strength. For instance, P24 noted that sometimes other family members step in to share caregiving responsibilities, giving them a needed break. 
\begin{quote}
\small
``[..] my brother comes [for caregiving], and we then can leave [..] My wife is very helpful, and she's able to be a little more [for caregiving tasks]''---P24
\end{quote}

% P15 mentioned that support from the care recipient themselves, when in good condition, also makes a positive difference. 
% \begin{quote}
% \small
% ``My mom always provide encouragement [..] when my grandpa (care recipient) is at least in a good condition, he he appreciate [what I am doing for him.]''---P15
% \end{quote}

% Similarly, P22 highlighted that when the care recipient is resistant to advice, having another family member step in is invaluable. 

% \begin{quote}
% \small
% ``Have my family come around to help, especially when it's proven difficult when doesn't want to listen to me at home.''---P22
% \end{quote}

% Family support ensures caregivers do not feel solely responsible, fostering a collaborative approach that is essential for sustaining their wellbeing.

\para{Social Support.}
Likewise, participants described that maintaining social life helps caregivers stay connected with the world beyond their caregiving responsibilities. 
% Engaging in casual or deeper conversations with family, friends, and professionals allows caregivers to shift focus away from caregiving and provides essential emotional support. 
For example, P11 noted, ``\textit{I have strong support from friends and neighbors, which helped immensely.}''
Similarly, P1, P21, P20, and P12 emphasized the importance of talking with friends, expressing that maintaining connections helps them process emotions and stay committed to their mental health journey. \edit{For example:}

\begin{quote}
\small
\edit{``Sometimes I just need to talk with my friends. It helps me feel grounded and not alone.''---P1}
\end{quote}

% \edit{
% \begin{quote}
% \small
% ``Now I make a point to communicate with family or friends when I need a break, whether that; 's having someone step in to care for a few hours or scheduling a respite care.'---P21
% \end{quote}
% }
% \edit{
% \begin{quote}
% \small
% ``It does provide some sort of support [..] sharing what I'm going through with others helps, and giving family and friends access [to my updates] means they can check on me and also collaborate with caring for my [care recipient]. It helps them understand what I'm going through and how to help me find balance.'---P12
% \end{quote}
% }
% \koustuv{Add a quote here.}
% }

% Our participant pool---which was largely recruited from Reddit---also highlighted how participating in online communities helps them engage in conversations with people experiencing similar issues.
% These platforms help them to self-disclose their concerns and also seek social support from other community members.
% Social interactions offer caregivers an opportunity for stress and serve as a reminder of life beyond caregiving, which is crucial for emotional resilience.

\para{Professional Support.}
% Professional help, such as counseling and therapy, is an important resource caregivers use to support their mental wellbeing. 
Several participants highlighted seeking professional mental health support such as therapy and counseling. 
For example, P17 mentioned that seeking therapy helped them control anxiety and the quality of caregiving.
In fact, P2 advocated for one-on-one therapy sessions rather than group therapy sessions, and P21 advocated for in-person therapies, that they tend to be more personal and connected:

\begin{quote}
\small
``In-person therapy: feels more personal and connected, physical presence creates a stronger send of trust, applauds the ability to share in a supportive, face-to-face setting.''---P21
\end{quote}

\para{Institutional Support.}
Institutional support from community, society, employers, and governmental agencies plays a critical role in helping caregivers manage the emotional and financial burdens of caregiving.
% , offering practical resources to alleviate stress
%For instance, P2 and P4 mentioned about how volunteer-led support groups also care about the caregivers' wellbeing and have connections with professional experts.
P10 emphasized the crucial role of employer support, particularly flexible work arrangements, in managing caregiving responsibilities.
Further, P10 mentioned a state program that levies taxes and provides stipends to help ease the financial burdens of caregiving:

\begin{quote}
\small
``There was a tax levied on all the residents of Connecticut several years ago, and now, starting in 2022, we can apply for a stipend based on our weekly or monthly income, which has been really helpful.''---P10
\end{quote}

% For example, P10 cited a state program that provides stipends based on income, which helped ease the financial strain. 
% P10 also mentioned 
% This kind of government assistance allows caregivers to focus more on their well-being while reducing some of the logistical pressures they face.
% \begin{quote}
% \small
% ``There was a tax levied on all the residents of Connecticut several years ago, and now, starting in 2022, we can apply for a stipend based on our weekly or monthly income, which has been really helpful.''---P10
% \end{quote}

\subsection{Caregivers Adopting Self-Care Practices}

Caregivers adopt a variety of self-care practices to maintain their wellbeing, including planning ahead, staying organized, exercising, engaging in hobbies, maintaining social connections, and setting boundaries. These practices help manage stress, preserve identity, and prevent burnout.
%. Participants emphasized that planning ahead and being organized helps them manage tasks and reduce stress. 
%They also noted that exercising and physical activities help rejuvenate themselves, and that hobbies offer a break from caregiving, preserving a sense of identity.
% Exercise and physical activities provide mental and physical rejuvenation, while hobbies offer a break from caregiving, preserving a sense of identity. 
% Participants also highlighted that maintaining social connections and setting boundaries between caregiving and social life helps them recharge and avoid burnout.
We further describe these themes below.
% Further, maintaining social connections prevents isolation and offers emotional support. 
% Setting boundaries between caregiving and social life ensures caregivers can recharge and avoid burnout.  

\para{Planning Ahead and Being Organized.}
Caregivers often find that staying organized and planning ahead reduces stress and promotes a sense of control in their caregiving responsibilities. 
% Structured routines, schedules, and pre-planned activities allow caregivers to manage both caregiving tasks and mental wellbeing more effectively. 
P6 mentioned that they use virtual personal assistants for calendar and reminders for time management, which help them to structure tasks, prevent overworking, and avoid mental strain:
\begin{quote}
\small
``[Virtual personal assistant will schedule task] in the calendar and I just get the reminders''---P6
\end{quote}

\para{Physical Activities and Breaks.}  
A majority of our participants highlighted the need for physical activities---spanning across simple outdoor walks (P1, P2), yoga (P4, P18), stretching and relaxations (P21), tai chi (P9), other exercise regimes (P15, P19). 
% Some participants (P2, P15) also mentioned that they include their care recipients in doing the physical activities together.
% to more structured exercise routines.
% , are essential tools for caregivers to manage their mental health. 
% Engaging in physical activities helps the caregivers to alleviate stress, provides emotional relief, and creates opportunities for caregivers to step away from their caregiving responsibilities. 
For example, P1 finds being outdoors therapeutic, as it allows them to connect with nature and experience a sense of peace.  
\begin{quote}
\small
``And I usually like being outdoors because I believe that when you're outside, you engage in the environment. You see beautiful things, beautiful nature, and then you can feel that. You can feel the, you know, the love of nature.''---P1
\end{quote}  

% Besides, P21 emphasizes the importance of taking short breaks when feeling overwhelmed, using physical movement like stretching or stepping outside to reset.

% \begin{quote}
% \small
% ``I've learned how to take short breaks when I feel overwhelmed. Even a few minutes of stepping outside or doing some stretching exercises can help me with it.'' ---P21
% \end{quote}  

% Exercise not only relieves physical tension but also supports emotional regulation, offering a necessary outlet for stress. By prioritizing movement, caregivers create time for themselves, disconnect from caregiving duties, and engage in self-care routines that promote mental well-being. Even brief moments of physical activity can reduce anxiety and improve mood, making it an effective mental health strategy.

\para{Hobbies.}  
Multiple participants also brought up the importance of pursuing hobbies to help distract their continued focus from caregiving tasks and stress.  
% Hobbies provide caregivers with a valuable mental break, helping them divert their focus away from caregiving tasks and alleviate psychological stress. 
% Each caregiver has unique hobbies that bring relaxation and joy. 
%We found a variety of hobbies that bring relaxation and joy to the caregivers.
%For example, P21 and P18 enjoy skincare routines as a way to unwind, P25 finds comfort in caring for pets, and P2 and P21 use journaling to process their thoughts. 
%Additionally, P18, P1, and P21 enjoy reading, while P5 and P18 do meditation. 
%Overall, we found that hobbies offer a mental escape, creating moments of joy, self-expression, and relaxation, which are essential for maintaining long-term emotional health and preventing caregiver burnout.
For instance, participants found comfort in skincare (P21, P18), pet care (P25), journaling (P2, P21), reading (P18, P1, P21), and meditation (P5, P18). These hobbies provided a mental escape, fostering joy, self-expression, and relaxation—key to sustaining emotional health and preventing burnout.

% \para{Maintaining Social Life}  
% \para{Maintaining Social Connections.} Participants described that maintaining social life helps caregivers stay connected with the world beyond their caregiving responsibilities. 
% Engaging in casual conversations or deeper connections with family, friends, and professionals allows caregivers to shift focus away from caregiving and provides essential emotional support. 
% For example, P11 noted, ``I have strong support from friends and neighbors, which helped immensely.''Similarly, P1, P21, P20, and P12 emphasized the importance of talking with friends, expressing that maintaining a social life helps them process emotions and stay committed to their mental health journey. Social interactions offer caregivers an opportunity for stress and serve as a reminder of life beyond caregiving, which is crucial for emotional resilience.

\para{Setting Boundaries Between Caregiving and Personal/Social Lives.}
%As previously noted, maintaining social connections emerged as a major theme for supporting wellbeing of caregivers. 
%However, p
Participants stressed the importance of setting boundaries between caregiving duties and personal or social life. %regardless of how entangled these aspects may become.
% Establishing boundaries between caregiving and other aspects of life is essential for preventing burnout. 
This included creating separate spaces (e.g., sleep arrangements, P9) and emotionally distancing during social interactions (P4, P17). As P17 shared:
%For example, participants brought up separating caregiving duties from personal space, such as separate sleep arrangements (P9) and emotionally distancing from caregiving during social interactions (P4, P17). 
%For example, P17 explained that they came to recognize the importance of self-care over the course of caregiving journey and began actively prioritizing it:

% Whether it’s physically separating caregiving duties from personal space (e.g., separate sleep arrangements) or emotionally distancing oneself from caregiving during social interactions, setting clear boundaries helps caregivers preserve their mental health.
% \begin{quote}
% \small
% ``I needed a little time on my own and I actually one of the ways I got this was I started sleeping separately from him.''---P9
% \end{quote}

% Setting boundaries allows caregivers to protect their personal wellbeing by creating necessary separation from the all-consuming nature of caregiving. 
\begin{quote}
\small
``[..] I've also become better at setting my own boundaries and prioritizing my own way of being early on. When I first started, I used to feel guilty about taking time for myself, but now I understand that I'm still essential for long term caregiving. So overall my experience has shifted from being taxed focused to a more balanced approach.''---P17
\end{quote}

%2 page
\section{RQ3: Technology to Support Caregivers' Mental Wellbeing}

Participants voiced the role of technology in both caregiving and their mental wellbeing. Technology serves as a multifaceted tools, offering practical helps for mental and emotional supports, as well as management in caregiving tasks. It enables caregivers to navigate complex needs while simultaneously focusing on self-care and work balance. 
% \koustuv{add another sentence about the overview of these technologies.}

\subsection{Need/Use of Technology}
Participants highlighted the need for technologies surrounding 1) caregiving responsibilities, 2) informational and learning resources, 3) communication and social connection, and 4) mental health and emotional support. 

% \subsubsection{Convenience of Tech}
% \subsubsection{Technology supporting caregiving responsibilities}

\para{Supporting caregiving responsibilities.}
Participants brought up aspects where they directly use different technologies to support their caregiving tasks.
These included cameras, smart locks, and tracking devices (e.g., AirTags) to monitor care recipients remotely, ensuring safety and easing emotional stress (P2, P13, P18, and P22). These tools helped manage risks such as wandering and accidents, especially when caregivers could not be physically present. For medication management and reminders, participants used smartphone reminders, Alexa, or automated dispensers like Hero to prevent missed doses and reduce caregiver stress (P12).
 
\begin{quote}
\small
``As she got into mild cognitive decline, I placed AirTags strategically in her purse, so I know where she is if we’re apart.''---P13
\end{quote}

\para{Information and Learning Resources.}
% Several participants mentioned how they often use a variety of technologies to get access to informational resources.
Several participants mentioned they often used technologies to access information.
They listened to podcasts (P1, P15, P22), watched YouTube videos (e.g., Dr. Natali Edmonds) (P13), and consulted sources like the Alzheimer’s Association website and Wikipedia (P5, P13). Online communities such as Reddit and Yahoo Groups provided peer advice and information on treatments (P2, P24). Some preferred short videos (P13), while others found AI tools like ChatGPT helpful for getting comprehensive, personalized answers (P20).

% P1, P13, P15, and P22 emphasized that they often listen to technology podcasts and Youtube videos to learn more about the AD/ADRD condition and caregiving. 
% For instance, P13 mentioned, ``I go look at YouTube videos, such as by Dr. Natali Edmonds---she has got a lot of caregiving stuff on YouTube.''
% \ravi{IIRC, the person I interviewed had a good quote about why he used youtube that we can add here.}
% P5 and P13 mentioned using the National Alzheimer's Association website and Wikipedia, respectively, to find reliable information. P13 specifically highlighted a preference for short video clips as a way to gather information.
% Multiple participants mentioned that they received information support and advice from online communities such as Reddit, where they can receive information from others facing similar challenges; P2 mentioned that Yahoo groups helped them connect with others as well as learn about clinical trials and various treatment procedures.
% Interestingly, P24 mentioned they only use Reddit asking questions about their mother's condition but avoids asking questions about their own mental health---``\textit{I don't know where I would go to ask.}''
% Some participants brought up their excitement with AI chatbots to receive detailed responses:

\begin{quote}
\small
``Unlike Reddit or Facebook, ChatGPT can help cover all my answers in one direction. I can even ask it for advice, and it even provides helpful suggestions on what I should do.''---P20
\end{quote}

% ---P20 mentioned that unlike Facebook or Reddit, ChatGPT can help cover their answers 

% brought up that social media and online communities also help them receive informational support and advice from online communities. 

\para{Communication and Social Connections.}
Participants used texts, emails, and video calls to stay connected with others and communicate with care recipients remotely. For example, P10 found Alexa Echo Show helpful due to its ease of use and support for lip-reading. Online communities also played a key role in reducing isolation and fostering a sense of belonging (P12, P13), especially when in-person socializing was limited.
%Participants noted the importance of technology in facilitating remote communication and maintaining social connections, such as using texts and emails to stay connected with friends, neighbors, and families. 
%In addition, they also use video calling technologies to contact their care recipient when remote---e.g., P10 mentioned found Alexa Echo Show to be helpful because it was easy to handle and their mother could lip-read because of suffering from hearing issues.
%In the context of social connectedness, here again, online communities play a vital part in ``feeling social'' especially when caregivers have limited time to socialize in-person. 
%For example, P12 especially noted the need to stay connected and reduce feelings of isolation during caregiving, and P13 noted that online communities make them realize, ``you aren't alone.''
%Several participants emphasized the ``sense of belonging'' provided by online communities when interacting with people with similar experiences.

\para{Mental Health and Emotional Support.}
To begin with, participants frequently mentioned how they would often receive emotional support from other community members in online communities---P11 mentioned that the Alzheimer's and Dementia Reddit communities were ``unusually kind for the Internet, being very much mutually supportive and kind.''

In addition, several participants use mental wellbeing apps, such as for meditation and relaxation (P11, P18, and P22), and for tracking wellbeing measures (P20 and P23). 
P1 uses the mental health app Headspace for mental health self-care, and also uses the app for personal journaling and emotional assistance.
Participants also continually tack their wellbeing through wearables and smart devices, such as Oura Ring (P10) and smartwatches (P2, P7, P9).
Tracking apps help caregivers monitor their own mental and physical health. 
% For example, P23 mentioned using apps like Ecom to track the emotional state and get motivation daily. 
% P20 highlighted an app’s functionality for tracking daily depression levels, offering a clearer understanding of mental health patterns. Similarly, P09 
These tools not only promote self-awareness but also encourage caregivers to prioritize their mental wellbeing, such as:
% P9 shared that they ``use a smartwatch and Fitbit to track steps and monitor health, providing motivation and encouragement through daily activity tracking.'' 
\begin{quote}
\small
``Even before Apple Watches, I had a Fitbit, and I counted steps on that Fitbit every day and looked at it to see what I’d done. It was motivating for me. These apps are motivating for me.''---P9
\end{quote}

Among more recent technologies, P14 mentioned that they use virtual reality (VR) based therapy and AI-based tools for stress relief. 
P2 noted that AI chatbots bear the potential to provide the emotional support and validation that caregivers often need.

\subsection{Perceptions about Technologies}
We now examine how our participants expressed varied perceptions of technology, which we categorize into---1) \textit{techno-optimism} and  2) \textit{techno-skepticism}.
% , and 3) perceived unhelpfulness.

% \koustuv{rename to techno-optimism and techno-skepticism}

\subsubsection{Techno-Optimism}

Given how technologies have been integral in their caregiving and mental wellbeing, several caregivers expressed optimism about the new and emerging technologies.
% Besides the fact that a majority of our participants had positive opinions about social media platforms, 
% Caregivers were optimistic about the potential of technology, particularly for caregiving and mental health management. 
As already noted, a majority of our participants were very positive about online communities.
P18 highlighted that virtual therapy sessions help them seek mental health services despite the time constraints.
In addition, speaking about other technologies such as apps, P1 noted that these function as ``another friend'' for a caregiver, and P15 particularly appreciated their 24/7 availability, allowing access to resources whenever needed.
Likewise, the timeliness and immediacy of responses that AI chatbots can offer were exciting to caregivers---for example, P20 described chatbots to be ``faster, convenient, detailed, versatile, and easy to access.''
Relatedly, P15 shared that, one night, his grandfather had an emergency, and it was impossible for him to get professional help at that time. 
They sought immediate response from web search and AI chatbot in getting guidance:

% 's immediate response provided important guidance, helping him navigate the situation effectively. These technologies offer caregivers quick, actionable support during emergencies when professional assistance may not be reaily available, ensuring timely interventions and reducing stress in critical moments. 
\begin{quote}
\small
``It was around 3:00 AM; I was worried about what was happening with my grandfather. You can’t call the doctor at that time, so I just googled and used an AI-based tool. I explained the symptoms and the condition, and the AI advised me to keep him active. We were able to resolve the situation. The key thing about [AI] is its availability---professionals are not always available.''---P15
\end{quote}

% Overall, several participants expressed their openness to trying new technologies that can ease caregiving or support mental health---P17 

% P1: the role of technology for caregiver is another friend, P15 appreciated the 24/7 availability that technology offers, allowing them to access tools and resources whenever needed. P20 mentioned AI-driven chatbot can provide faster, convenient help. P18: Reddit, where they could connect with others facing similar caregiving challenges and mental health situation. P18: see her therapist online. 

\subsubsection{Techno-Skepticism}
Despite the positive outlook highlighted above, caregivers also expressed varying degrees of skepticism about technology in caregiving and mental health context. 
P17 noted that despite the notable usefulness of technologies, it is important to approach them with caution, prioritizing reliability and privacy protection.
Several participants advocated for real human interactions over interactions with an AI, like P24 labeled an AI chatbot as a ``data-center.''
P20 added that although Reddit is usually helpful, it does not enable face-to-face communication.
% , which they would prefer having.
Particularly about emerging AI chatbots, a common concern was their reliability and trustworthiness in providing medication advices.
% .medication and emotional support. 
P2 emphasized the need for credible sources in responses by AI chatbots. 
Additionally, regarding emotional support, some caregivers expressed that AI lacks the ability to understand complex human emotions and medical information:

\begin{quote}
\small
``I'm concerned about something like ChatGPT, how it would understand my emotions and provide accurate advice based on my feelings.''---P12
\end{quote}

% Despite these concerns, caregivers acknowledged the potential benefits of AI and other emerging technologies. They saw future possibilities where AI could be more refined and helpful. 

% \subsubsection{Perceived Unhelpfulness}

% Some caregivers found technology to be unhelpful or frustrating in certain contexts. P11 is not an app user because things were never predictable enough to be able to concentrate fully on using an app. 
% Participants voiced that the data-driven technologies lack accuracy; for example, P5 was unhappy with the frequent interventions and notifications by their smartwatches, and P3 found inaccuracies in ChatGPT's responses.
% P2 found virtual therapy to be ineffective on them.
% Additionally, one of the main criticisms was the lack of emotional connection and personalization---not being able to account for specific needs.
% % Some participants also found AI chatbots to be too generic and not able to account for specific needs.
% % , the solution that didn't account for the specific needs of individual caregivers. 
% The lack of personalizations left caregivers feeling as though the technology was not truly supporting them in a meaningful way.

\subsection{Challenges and Proposed Improvements}\label{sec:tech_challenges}
% \melissa{the table not accurate yet, need to update}

Finally, we examine the technology-related challenges faced by the participants and improvement recommendations that emerged from our interviews.~\autoref{tab:challenges-improvements} provides an overview of mapping the challenges and proposed improvements, and we elaborate on this below.

\begin{table}[t]
\centering
\footnotesize
\sffamily
\caption{Challenges and proposed improvements for technologies to support mental wellbeing of caregivers.}
% \begin{tabular}{>{\columncolor{white}}p{0.3\linewidth}>{\columncolor{white}}p{0.6\linewidth}}
\begin{tabular}{p{0.25\columnwidth}p{0.7\columnwidth}}
\textbf{Challenges} & \textbf{Proposed Improvements} \\
\toprule
% \midrule
Financial Barriers & Offer more affordable pricing models, such as sliding-scale subscriptions or free basic version of apps, to improve accessibility. \\
\rowcollight Technical Complexities & Design user-friendly interfaces with simplified navigation, reduce notifications, minimize complexity and streamline key features. \\
Data Security and Concerns & Enhance safe data storage and transparency in data usage policies. \\
\rowcollight Reliability and Credibility of & Incorporate verified, professional information to build trustworthiness and ensure accuracy for decision-making. \\
Incorporating Human Interaction & Integrate human interaction features, connecting with other caregivers in a non-anonymous format. \\
\rowcollight Lack of Personalization & Provide personalized support through customized medical care recommendations and individualized content to meet specific caregiver needs.  \\
\bottomrule
\end{tabular}
\label{tab:challenges-improvements}
\end{table}

\subsubsection{Financial Barriers}
% \para{Challenge}
Many caregivers reported financial barriers when using caregiving technologies, particularly subscription fees for some apps. 
For example, P15 mentioned the cost was a significant obstacle preventing them from fully utilizing the technology.
% , noting, \textit{"If I didn't pay for some AI-based chatbot, it has limited times to ask questions about caregiving."} 
Similarly, P18 discontinued the use of an app to improve sleep beyond its free trial period.
% tried an app designed to improve sleep but discontinued its use due to the need for payment after the trial period:
% \begin{quote}
% \small
% ``I tried an app to help me sleep. But they wanted you to pay [..] so I didn't like that.''---P18
% \end{quote}
% P17: high costs of premium versions
% \begin{quote}
% \small
% ``The cost is actually one of the issues [...] because you were to manage the cost. Public subscription. Some kind of mental health apps.''---P20
% \end{quote}
% \para{Proposed Improvement}
Consequently, caregivers suggested more affordable pricing models, such as sliding-scale subscriptions or free basic versions of apps, enabling broader access across various income levels and reducing financial strain on caregivers.

\subsubsection{Technical Complexities}

% \para{Challenge.}
% \melissa{Add time constraints for one of challenge, since spend a lot of time on how to figure out}
Technical complexity emerged as a major issue for caregivers, particularly those who may not be tech-savvy. 
P2 noted that there was no customer support to help them navigate through the features of apps. 
% P17 was concerned about how their lack of technical skills, further complicated how they used the softwares and tools.
Interestingly, P9 noted that apps are becoming increasingly complex with added security features, making it hard to keep up, especially at their age:

\begin{quote}
\small
``[Technologies are] getting so complicated for me. [..] Because of all the Internet scams, the security levels are going up everywhere, and this is very challenging to keep up with.''---P9

\end{quote}

% P2: hard to find the right person to talk with since difficult for navigate on apps. For instance, P2 mentioned that it was hard to find the right person to talk with due to difficulties navigating apps. Similarly, P17 highlighted issues with software and a lack of technical skills, further complicating their use of these tools. P9 noted that apps can be complicated and challenging, particularly with the rise of security concerns and online scams. This inaccessibility, with technical challenges, becomes a major barrier for caregivers trying to utilize technology effectively. 
% \begin{quote}
% \small
% ``[...]so they should make the app easy to use, not complex, which makes the user interface very very good, so the user is able to do what they can do at the right time.''---P20
% \end{quote}
% \begin{quote}
% \small
% ``Maybe some kind of more efficient interface to Reddit. Now for Reddit, you can sort by best or by newest, or by most active, but the search mechanisms there in Reddit aren't that great, and so if I'd had. Experiences that really were unusual and I needed to find out who else has experienced this particular situation.''---P11
% \end{quote}
% \para{Proposed Improvement}

To overcome the technical barriers, caregivers emphasized the need for more user-friendly interfaces with simplified navigation that reduces the time spent figuring out how to use tools. 
For instance, P4 suggested that apps should avoid excessive notifications since caregivers have already been overwhelmed by caregiving tasks. 
P11 added that it is an efficient interface, especially when searching for specific caregiver experiences or advice. 
Importantly, caregivers---already pressed for time---need technology that is more accessible, especially as many are older adults with limited digital literacy.
By minimizing complexity and streamlining key features, caregiving technology can become more accessible and useful for a wider range of caregivers.

\begin{quote}
\small
``[..] so they should make the app easy to use, not complex, which makes the user interface very good, so the user is able to do what they can do at the right time.''---P20
\end{quote}

\subsubsection{Data Security and Concerns}

First, multiple participants expressed that they are not concerned about data security and privacy on the technologies.
Their lack of concern stemmed from two reasons: (1) they avoid posting personal or sensitive information online when seeking help, and (2) they are already de-sensitized to the pervasiveness of data tracking, as highlighted P18's comment, ``this is the internet, and everyone’s tracking everything.''
However, some participants expressed concerns---P20 was concerned about the uncertainty on how their data on the Internet might be used, and P12 was concerned about privacy and security on caregiving apps, emphasizing the need for safe data storage to prevent leaks or exposure on social media.

% primarily due to uncertainty about how their data might be used (P. 

% \melissa{Question about this theme, since 50\% participants: no data security concern, since(1) they avoid post personal or sensitive information online when they ask for help; (2) this is internet and everybody's tracking everything--P18, but 50\% has data concern about no idea the data will be used. So, Im thinking do we need keep this theme?}

% \para{Challenge}

% \para{Proposed Improvement}

\subsubsection{Reliability and Credibility of Information}

Caregivers expressed concerns about the accuracy and credibility of information from the internet or AI, especially when using unverified sources. These worries relate to health and safety, as caregivers need reliable support for decision-making. P20 noted that online searches can sometimes yield outdated or inaccurate guidance, while P2 and P3 emphasized the importance of trustworthy sources for effective caregiving. P17 added, ``I'd like a feature that allows you to speak with a professional.''

% \para{Challenge}
% Caregivers expressed concern about the accuracy and credibility of the information they receive from the Internet or AI.
% % , especially when the they reply on AI or unverified sources. 
% Caregivers pointed out their health and safety concerns: worry about effectiveness, decision-making. For instance, P20 noted that online search responses sometime are unreliable and do not always provide the most accurate or up-to-date guidance. P2 acknowledged the potential benefits of technology for caregivers but stressed the importance of ensuring that the information comes from credible and trustworthy sources. Similarly, P3 emphasized the need for tools that a trustworthy system is vital for managing caregiving tasks effectively. P17 "I'd like a feature that allows you to speak with a professional. "
% \begin{quote}
% \small
% ``[...]would like from a technology is trustworthiness. It needs to give accurate information I am looking for both like information related to mental well being about your caregiving.''--P03
% \end{quote}
% \para{Proposed Improvement}
To address these concerns, caregivers proposed that caregiving technologies should prioritize trustworthiness by incorporating information that is verified by professionals. 
Providing caregivers with transparency regarding where the information comes from, along with the verification process, would further build trust. 
In doing so, caregivers could rely on technology for decision-making, knowing that it has been informed by credible sources.

\subsubsection{Incorporating Human Interaction}
% \para{Challenge}
A key challenge that caregivers face when using an automated technology is the lack of human interaction, which is crucial in providing emotional support. 
For example, P12 and P14 sought the ability to communicate with healthcare professionals through the technology, rather than just relying on automated responses or pre-programmed features. 
P8 emphasized the need for a human interaction interface to help navigate the emotional and complex decisions that arise during caregiving. 
% P20 mentioned empathic; P1: 
P1 sought more space where caregivers could interact with each other. 
Without these interactive, human-centric features, caregivers may feel unsupported when they need advice or validation, particularly in high-stress or crisis moments: 
% Additionally, the emotional aspect of caregiving often requires human empathy-something that current technologies struggle to replicate. 
\begin{quote}
\small
``Emotions are something that requires human interaction than technology to deal with''--P8
\end{quote}

% \begin{quote}
% \small
% ``[...]So that would be nice if you had an app like where you know where everybody is and not just anonymously. Not like Reddit, you know, it's all anonymous. It would be nice to have, uh, like in person or even like, you know, FaceTime meetings or whatever or people who are just like me.
% ''---P18
% \end{quote} 

% \para{Proposed Improvement}
To address this concern, caregivers suggested integrating more human interaction elements into caregiving technologies. 
For instance, P14 mentioned that the VR therapy also offers a space to talk to an expert anonymously.
P18 sought for a platform where they can meet with other caregivers in a non-anonymous fashion (unlike Reddit):

\begin{quote}
\small
``[..] So that would be nice if you had an app where you know where everybody is and not just anonymously. Not like Reddit, it's all anonymous. It would be nice to have in-person or even FaceTime meetings or whatever with people who are just like me.
''---P18
\end{quote} 

% : Virtual reality therapy offers a space to talk to someone anonymous. P18: she wants to have a technology where she can meet everyone in ``non-anonymous'' fashion (unlike Reddit) and also have FaceTime meetings. The proposed improvements center around enhancing the human aspect of caregiving technology, ensuring that while technology provides efficiency and convenience. 

\subsubsection{Lack of Personalization}
% \para{Challenge}
Several caregivers felt that existing technologies offers generic help that fail to cater to their specific contexts and unique needs. 
For instance, P5 mentioned that caregivers' needs are unique, and there is no one-size-fits-all solution. 
% and there is no one-size-fits-all solutions. Current support systems fail to address the diverse challenges faced by AD/ADRD caregivers. 
% \begin{quote}
% \small
% ``I think the big hurdle with it is every caregiver is different and the needs are different''---P5
% \end{quote}
% \begin{quote}
% \small
% ``I would like personalization [...] I really like to that are able to adapt to each clients unique need and behavior.''---P17
% \end{quote}
% \para{Proposed Improvement}
The participants
% refore, caregivers 
proposed personalization in caregiving technology. They suggested that apps could provide personalized medical care, including tailored medical recommendations (P14).
P5 desires for a trusted personalized source of information and support that also emphasizes shared experiences and individual expressions over generic solutions.
% \melissa{Can we skip the theme: MH intervention, I feel the improvement proposed in this theme, a lot of tech, app already has the functionality. }
% \subsubsection{Lack of Information Support}

% \para{Challenge}
% In addition to the challenges mentioned above, caregivers highlighted gaps in other essential functionalities within caregiving technologies. Many caregivers noted that they need more comprehensive \textit{informational support} and \textit{mental health interventions} integrated into the tools they use.Current platforms often lack up-to-date, easily accessible information on caregiving best practices or resources specific to different caregiving stages. 

% \para{Proposed Improvement}. 
% For example, P6: tools can provide educational resources; P13: tools offer clear guidance on AD/ADRD's stages and prefers practical advice over drug updates; P11: tools could pull together all available resources for aging, AD/ADRD support, that could standardize data from different facilties and show real-time availability, 
 %2 page
% \input{3taxonomies} %2 page
% \input{4recruitment} %2 page
% \input{5analyses} %2 page
% \input{6triangulation} %2 page
% \input{7discussion} %2 page
\section{Discussion}\label{sec:discussion}

In this study, we conducted semi-structured interviews with 25 family caregivers of individuals with AD/ADRD, and adopted inductive coding and thematic analyses~\cite{braun2019reflecting} to identify the 1) major mental health concerns of caregivers (RQ1), 2) practices and coping strategies employed by caregivers (RQ2), and 3) technologies used by caregivers in their daily caregiving and support mental wellbeing. 
% Our study explored the mental wellbeing needs of 25 family caregivers supporting individuals with AD/ADRD. 
At a high level, the interviews reinforced the motivation of our study, and confirmed a critical gap in the current healthcare infrastructure and support system---caregivers' needs are largely unaddressed, with few systematic mechanisms tailored to prioritize their mental health. 
In particular, AD/ADRD predominantly affects elderly individuals, whose caregiving responsibilities typically fall into one of two age groups---1) middle-aged adults caught in the ``sandwich generation,'' simultaneously needing to caregiving for aging parents while parenting their children, or 2) aging spouses who face their own health challenges---``I don't want to die before my husband''---as P9 emotionally reflected about a deep-concern for their future.
In this section, we discuss the implications for our work. 

\subsection{Theoretical Implications}
\edit{In this section, we discuss how our findings are situated with existing theoretical frameworks.
In particular, we contextualize our work with two relevant theoretical lenses---Social Support Behavioral Code (SSBC)~\cite{suhr2004social} and the Ethics of Care~\cite{gilligan2014moral,tronto2020moral}.}

\subsubsection{Caregivers need and seek social support}
The need for social support emerged as a major theme in protecting the wellbeing of caregivers, and underscores the necessity for sociotechnical systems that support their needs.
We situate these observations with the Social Support Behavioral Code (SSBC)~\cite{suhr2004social}, which categorizes support into five types---informational, emotional, esteem, tangible, and social network support~\cite{suhr2004social}. 
Our findings demonstrate that caregivers require and seek these forms of support to cope with their mental wellbeing concerns.

\para{Informational Support.} Our participants emphasized the need for reliable advice, information, and suggestions to enhance their caregiving practices and manage their mental health. 
They frequently turned to podcasts, YouTube channels, and reputable health websites. 
Prior work in this space by~\citet{wong2024voice}, similarly noted that caregivers value technologies like Voice Interface Personal Assistants (VIPAs) that provide informative content to support older adults' mental health.

\para{Emotional Support.} Emotional burden was a recurring theme, with participants reporting anxiety, depression, and emotional exhaustion. Emotional support is highly contextual; the type of support an AD/ADRD caregiver seeks may differ significantly from that of others.
Caregivers sought emotional support through professional counseling, therapy, and maintaining social connections with friends and family. This need for emotional solace is echoed in~\citet{siddiqui2023exploring}, where caregivers of people with serious mental illness benefited from peer support groups that offered empathy and understanding, and~\citet{kim2024opportunities} highlighted the importance of emotional support for informal dementia caregivers dealing with verbal agitation.

\para{Esteem Support.} Our findings revealed that caregivers often felt unappreciated and overwhelmed, leading to diminished self-worth. The lack of positive feedback or acknowledgment of their caregiving efforts contributed to feelings of ``thanklessness.'' 
We noted how participants felt they would feel much better when their loved ones would appreciate their efforts. 
Similarly, in social interactions, positive messages that reinforce their esteem can help bolster a caregiver's self-worth and resilience.
Participants self-reflected a positive outcome from their caregiving journey, in terms of personal growth and enhanced self-worth---this aligns with~\citet{meyerhoff2022meeting} which revealed how recognizing caregivers' efforts can enhance their sense of value and reduce feelings of isolation.

\para{Tangible Support.} Physical assistance and provision of goods or services were vital in alleviating caregivers' burdens. 
Some participants relied on government programs for financial aid and on family members to share caregiving tasks. 
However, several reported that this support was insufficient and expensive, leading to heightened stress and financial strain. 
This observation aligns with~\citet{siddiqui2023exploring}, which found that inadequate healthcare support systems place an extreme burden on caregivers, emphasizing the need for more robust tangible support mechanisms.

\para{Social Network Support.} Caregivers expressed a strong desire to feel connected to a community of individuals facing similar challenges---both offline and online. 
Online platforms, particularly Reddit, provided a sense of belonging and facilitated the exchange of experiences and advice~\cite{johnson2022s,levonian2021patterns,boessen2017online}. 
This helped mitigate feelings of isolation and fostered emotional resilience. 
This aligns with a large body of prior work in social computing and social support, highlighting the therapeutic effects of self-disclosure and social support in online communities~\cite{saha2020causal,ernala2017linguistic,andalibi2017sensitive,de2014mental,kim2023supporters}, and more specifically~\citet{foong2024designing} emphasized the importance of social connections in supporting caregivers' mental health, suggesting that community engagement can be a valuable resource.

\subsubsection{Situating with the Ethics of Care}
We also situate our findings with the Ethics of Care framework~\cite{gilligan2014moral,tronto2020moral}.
This framework emphasizes the moral significance of relationships and dependencies in human life, highlighting the importance of caring practices and the wellbeing of both caregivers and care receivers. Our findings reflect key principles of this framework:

\para{Relational Tensions and Emotional Labor.} Caregivers face significant challenges in managing family relationships and tensions, often feeling overwhelmed by caregiving responsibilities and lack of family support. 
The emotional labor involved in caregiving led to additional strain. 
This supports prior work~\cite{meyerhoff2022meeting}, and with the Ethics of Care~\cite{gilligan2014moral,tronto2020moral} emphasis on the complexities of care relationships and the moral importance of attending to these relational dynamics~\cite{gilligan2014moral,tronto2020moral}. 

\para{Dependency and Vulnerability.} 
The progressive decline of care recipients' health heightened caregivers' emotional stress, especially when care recipients lost memory or recognition of the caregiver. 
Caregivers experienced anticipatory grief and a sense of impending loss. 
The Ethics of Care framework~\cite{gilligan2014moral,tronto2020moral} highlights the moral significance of responding to dependency and vulnerability, advocating for support systems that address these challenges.

\para{Maintaining Caregivers' Wellbeing.} The Ethics of Care calls for promoting the wellbeing of caregivers within a network of social relations. Our findings show that caregivers' own wellbeing was compromised due to financial strain, social isolation, and lack of self-care. Technologies and support systems need to consider the caregivers' needs, not just those of the care recipients---aligning with~\cite {kim2024opportunities}'s emphasis on designing technologies that support informal caregivers' mental health during unpredictable verbal agitation from people with dementia.

\subsection{Societal and Policy Implications}

Our study bear implications for societal structures and policy-making, particularly in relation to financial, social, and mental health support for AD/ADRD caregivers. These insights highlight both systemic gaps and opportunities for multi-level interventions that prioritize caregiver wellbeing.
\edit{We preface this discussion by acknowledging that our study was conducted in the United States and as such the topics and responses reflect the socio-economic contexts of the region.}

\subsubsection{\edit{Financial Support as a Foundational Need}}
\edit{One of the most pressing mental health concerns reported by caregivers was the significant financial strain associated with long-term caregiving responsibilities.}
% To begin with, the financial strain experienced by caregivers points to the need for comprehensive support systems. 
Policies that provide financial assistance, such as stipends or tax credits, can alleviate economic burdens and reduce anxiety related to the sustainability of care~\cite{zarit1986subjective, kim2024opportunities}.
\edit{These recommendations align with the World Health Organization (WHO)'s Global Action Plan on Dementia, which calls for financial support systems for caregivers~\cite{who2017global}, and Alzheimer Europe's position paper advocating for economic relief through stipends and tax credits~\cite{alzheimereuropeDementiaEurope}.}
Interestingly, multiple participants highlighted that, although institutional and healthcare support exists for patients, there is a severe lack of similar resources dedicated specifically to caregivers\edit{---highlighting a gap between policy recommendations and implementation that is also noted in the U.S. National Plan to Alzheimer's Disease~\cite{NationalPlan2025}. 
\edit{The 2018 RAISE (Recognize, Assist, Include, Support, and Engage) Family Caregivers Act is a step in this direction~\citet{acl_raise_council}.}
Our findings point to a need for future work that not only advocates for financial policy change but also  systems that help caregivers navigate financial aid, manage caregiving-related expenses, or coordinate financial planning across family networks. 
Particularly, policies need to lower the barriers to accessing the latest advances in technology and medicine---from speeding up regulatory approval for new products and practices to negotiating better coverage through insurance. 
}

% Further, the disrupted social life and social isolation reported by caregivers highlight the importance of community-based interventions. 
\subsubsection{\edit{Combating Social Isolation Through Community-Based Interventions}} We found that caregivers often face social isolation and disrupted social lives, underscoring need for community-based interventions that go beyond traditional support groups to foster meaningful connections and practical support.
% These interventions can build on and extend traditional support groups to create meaningful social connections and practical support networks.
Local community centers, faith-based organizations, and neighborhood support programs could offer respite care services, organize social activities that accommodate caregivers' scheduling constraints, and facilitate peer-to-peer connections among families facing similar challenges~\cite{dai2023library}.
For instance, P2 shared that their church barred their partner from attending services after AD/ADRD progression---significantly worsening the social lives of both of them. \edit{This echoes for~\citet{ruitenburg2024evolving}'s call for policy-level interventions aimed at reducing the stigma associated with dementia-related stigma and inclusion in key social spaces. 
% communication challenges and promoting more inclusive social environments.
% The current anti-stigma efforts may not sufficiently engage influential social settings such as faith-based institutions.
As faith-based institutions and churches often serve as vital community hubs---especially in rural areas---exclusion can lead to broader social ostracism.}
% This is a particularly critical problem to address given that church is the most common third space in many parts, especially in rural regions, of the U.S. and being ostracized from one's church can be akin to being ostracized from the entire local community.}
% Essentially, there is a critical need for greater awareness of the condition and how social support and acceptance can help both the care recipient and the caregiver.

% Programs that facilitate social connections, such as caregiver support groups and respite care services, can help mitigate loneliness and emotional distress. 
To address this gap, public awareness campaigns can de-stigmatize caregiving challenges and promote community support~\cite{dai2023library, kim2024opportunities}. \edit{While Alzheimer Europe's anti-stigma initiatives offer a foundation~\cite{alzheimereuropeDementiaEurope}, our participants' experiences indicate that current approaches may be insufficiently reaching faith-based and community organizations. Therefore, our study recommends the need for more localized, context-sensitive engagement strategies---especially within community organizations that caregivers frequently interact with. This extends~\citet{nunes2015self}'s recommendations by emphasizing social, not just clinical, pathways to support.}

\subsubsection{\edit{Improving Access to Mental Health Support for Caregivers}} 
\edit{In addition to financial and social challenges, our findings point to two critical systemic gaps that impact caregiver wellbeing---1) the limited availability of caregiver-specific mental health services, and 2) the administrative and institutional burden of accessing support systems. 
Many participants reported unique psychological stressors---such as emotional upheaval~\cite{blum2010family}, compassion fatigue~\cite{day2011compassion}, anticipatory grief~\cite{millenaar2018exploring,vaingankar2013perceived,wawrziczny2017spouse}, and a sense of hopelessness about the future---that extend beyond traditional diagnoses like depression and anxiety~\cite{schulz2008physical}. These findings underscore the need for tailored mental health services that account for the caregiving context~\cite{randall2018engaging, lederman2019support}.
Although the recent changes to U.S. medicare allow caregivers to bill for their services, it is still not comprehensive to cover their health needs stemming from caregiving work~\cite{actonraise2024pfs}. 
The RAISE act can be a potential solution if it is applied to cover the above needs of caregivers~\citet{acl_raise_council}.
Towards supporting the administrative burden, we propose leveraging the Digital Navigator Model~\cite{perret2023standardising,ndia2024digitalnavigator}. 
Essentially, providing caregivers and patients access to navigators/community workers who can help them navigate the administrative aspects of caregiving --- accessing records, finding relevant government support programs, connecting with providers, finding respite. 
}

\edit{To address these needs, accessible and proactive mental health interventions are essential, such as counseling, peer support programs, and targeted resources for caregivers.
Embedding mental health screenings into regular healthcare visits can help detect distresses early and prevent crises~\cite {kokorelias2022grounded, dayer2013smartphone}.
% a preventive approach to identifying caregivers' distress early and intervening before crisis points~\cite{kokorelias2019p2, dayer2013smartphone}.
At the same time, caregivers face significant friction in accessing benefits and services due to bureaucratic complexity. Participants noted how time-consuming processes detract from caregiving. 
Streamlining access to public programs through simplified applications, better guidance, and integrated platforms can ease this burden~\cite{kim2024opportunities}. This requires collaboration across healthcare providers, social services, and community organizations~\cite{kim2024opportunities}---a system-level approach that positions caregiver wellbeing as a central, not peripheral, concern.}
% This burden could be reduced by streamlining access to public programs through simplified applications, better navigational assistance, and integrated platforms~\cite{kim2024opportunities}. 
% This requires collaboration across healthcare providers, social services, and community organizations~\cite{kim2024opportunities}---a system-level approach that positions caregiver wellbeing as a central, not peripheral, concern.}

\subsection{Technology and Design Implications}
\edit{Our research extends prior HCI and CSCW work on AD/ADRD caregiving technologies, such as those supporting care transitions~\cite{houben2024design}, robotic assistance in daily care~\cite{lee2023reimagining}, and online peer support communities~\cite{johnson2022s}, by centering on the \textit{mental health needs} of caregivers and how these needs evolve throughout the caregiving journey. 
We draw on the concept of evolving interpersonal dynamics highlighted by~\cite{ruitenburg2024evolving}, which explores how dementia disrupts relational communication and affects caregivers' sense of emotional connection. 
While prior research suggested task-oriented or functional support systems, our study foreground caregivers' perceived challenges and the need for technologies that also address emotional resilience, anticipatory grief, isolation, and feelings of guilt related to self-care. Rather than solely facilitating caregiving tasks, our design implications call for systems that enable self-reflection, coping strategies, and emotional validation---particularly during transitional and high-stress periods of caregiving experience.}
% highlights the family caregivers' perceived concerns and desire for technologies that can also address emotional resilience, anticipatory grief, and feelings of guilt (about self-care) and isolation. In contrast to primarily task-oriented tools, our implications suggest a need for systems that support self-reflection, coping practices, and emotional validation—especially during key transition periods.}
% \edit{We found that caregivers' use of technology varies across different needs---from monitoring~\cite{boessen2017online} and medication management to social belonging and emotional support.}
% Our findings suggest that caregivers' use of technology varies across different needs---from monitoring and medication management to social belonging and emotional support. 
Although caregivers rely heavily on digital tools for practical tasks, their emotional needs for human interaction, personalized support, and credibility are not always adequately met by current technological solutions. 
\edit{Along these lines, we first, highlight some of the cross-cutting needs for caregiver wellbeing technologies, followed by three major design implications from our work.}

% Our findings suggest several technological and design implications for improving the mental health and caregiving experience for AD/ADRD caregivers. 
% \edit{Our findings suggest several technological and design implications directly tied to our results on caregivers' experiences and needs.}

\subsubsection{\edit{Cross-cutting Needs for Caregiver Wellbeing Technologies}}
\edit{Aligning with a significant body of prior work~\cite{schorch2016designing,chen2013caring}, our findings highlight three major cross-cutting needs that must be addressed to ensure technologies effectively support caergivers' mental wellbeing, as described below:}

\para{\edit{Affordability and Accesibility.}} \edit{First, affordability remains a critical barrier. Participants like P15 and P18 reported abandoning tools due to cost, even when these tools were perceived as helpful (\autoref{sec:tech_challenges}). Financial strain was especially pronounced in the early stages of caregiving (\autoref{sec:rq1}), reinforcing the need for low-cost, subsidized, or open-source mental health technologies~\cite{foong2024designing, lucas2022experience}.}

\para{\edit{Usability and Personalization.}} \edit{Second, usability challenges can discourage adoption, especially among caregivers managing emotional fatigue. P9 noted frustration with increasingly complex digital systems, exacerbated by security changes (\autoref{sec:tech_challenges}). Multiple participants expressed interest in tools like Alexa that provide simple and desired features (\autoref{evolution})---aligning with prior findings~\cite{zubatiy2021empowering}. Personalized, adaptive systems---such as mood-tracking tools---could offer timely wellbeing interventions~\cite{yamashita2017changing}.}

\para{\edit{Credibility.}} \edit{Finally, caregivers raised concerns about the credibility of digital guidance, particularly in emotionally complex situations. P2 and P12 questioned the ability of AI tools like ChatGPT~\cite{cgpt} to provide emotionally accurate or trustworthy advice. Prior work has shown that involving healthcare professionals in content development can help ensure both clinical accuracy and emotional relevance~\cite{de2023ethical,sharma2024facilitating}. Additionally, implementing a ``source transparency'' framework---where tools clearly disclose the origins of their recommendations (e.g., medical literature, professional guidelines, or clinical best practices)---can further enhance trust~\cite{ehsan2023charting,quinnmobile,yoo2025patientcentered}. Such transparency has been shown to improve perceived reliability in digital health systems~\cite{kelley2024mobile}. Our findings reinforce that credibility is not just a technical issue---it is central to emotional reassurance, cognitive ease, and sustained engagement, especially as caregivers navigate high-stress and uncertain care contexts.}

% Together, these needs point to the importance of designing technologies that are not only accessible and usable, but also emotionally attuned and contextually credible.

% \koustuv{Shrink and add about accessibility, user-friendliness, and credibility.}

\subsubsection{\edit{Designing for Emotionally Aware Caregiving Technologies}} \edit{Building on caregivers' evolving mental health needs, our findings point to several promising directions for emotionally supportive technologies. For instance, we can think of designing \textbf{anticipatory emotional guidance tools} that offer context-sensitive reflections during major care transitions—layering psychological support onto existing logistical features~\cite{houben2024design}.
In addition, building on identity-aware design from prior work~\cite{ruitenburg2024evolving}, we can design \textbf{self-reflective interfaces}, which can help caregivers better recognize and articulate emotional strain or feelings of relational distance---as well as help with the personal growth and emotional maturity (as noticed a positive mental health effect of caregiving). 
We found several instances of referring to the online communities for emotional and informational support. This also support a series of prior work on the potential benefits of online support communities, both generally~\cite{andalibi2016understanding,saha2020omhc}, as well as in the case of AD/ADRD caregivers~\cite{kaliappan2025online,johnson2022s}. 
Our findings encourage design discussions on building peer support platforms that can also facilitate co-regulation of emotions, shared coping strategies, and the normalization of difficult feelings such as guilt, burnout, or grief. 
Finally, emotion-aware or mood-sensing interfaces can complement task-based assistance~\cite{lee2023reimagining} by engaging with caregivers' internal states---offering low-effort emotional check-ins or personalized nudges aligned with caregivers' energy levels and caregiving intensity---recent research integrated behavioral sensing and generative AI capabilities for adaptive mood interventions~\cite{das2025ai,nepal2024mindscape}. These directions highlight the need to treat mental wellbeing not as an add-on, but as a central component of caregiver technology design.}

\subsubsection{Integrating \edit{Automated} Technologies with Human Interactions}
Caregivers \edit{in our study consistently} reflected on the emotional burden of caregiving, with many experiencing anxiety, depression, and emotional exhaustion. 
Emotional support is highly contextual; the type of support an AD/ADRD caregiver seeks may differ significantly from that of others.
Caregivers sought emotional support through professional counseling, therapy, and maintaining social connections with friends and family. This need for emotional solace is echoed in~\citet{siddiqui2023exploring}, where caregivers of people with serious mental illness benefited from peer support groups that offered empathy and understanding, and~\citet{kim2024opportunities} highlighted the importance of emotional support for informal dementia caregivers dealing with verbal agitation.
\edit{In addition, our findings highlighted caregivers' desire for a more hybrid approach---combining human connections with technological scaffolding.}

In particular, the participants noted the lack of empathetic interaction in current automated technologies. \edit{P8 stated, ``Emotions are something that requires human interaction than technology to deal with.'' This reflects our findings on compassion fatigue and emotional upheavals (\autoref{sec:concerns}), where caregivers struggle with complicated emotions that purely automated solutions cannot adequately address.} 
Integrating features that facilitate connections with professionals or peer caregivers can provide emotional support. 
For example, platforms that offer virtual support groups or telehealth consultations can bridge this gap. \edit{Several participants, including P14, expressed interest in VR therapy that ``offers a space to talk to an expert anonymously,'' while P18 sought ``a platform where they can meet with other caregivers in a non-anonymous fashion.'' These preferences directly connect to our findings on social support seeking (\autoref{sec:practices}), where connecting with others facing similar challenges fostered a sense of belonging, mutual understanding, and emotional solidarity—key elements in building emotional resilience for caregivers.}

Additionally, \edit{caregiving is multi-layered and collaborative}; caregivers desire technologies facilitating collaboration with other caregivers and stakeholders (e.g., coordination with family, healthcare providers, and community resources). 
\edit{For this, technologies can incorporate features such as shared calendars, task delegation tools, and secure communication channels. Existing platforms such as CareZone or Lotsa Helping Hands address some of these needs, but our findings suggest extending these to integrate emotional wellbeing tracking~\cite{yamashita2017changing}, mutual check-ins, and caregiver burnout alerts---bringing emotional support into shared caregiving infrastructures.}

\subsubsection{\edit{Dynamic Designs: The Future of Technology to Support Caregivers' Mental Health}}

\edit{Prior work has advocated for designing technologies that respect the integrality of caregiving---viewing caregivers not merely as support agents for others, but as individuals with complex, intersecting emotional and logistical needs~\citet{chen2013caring}. 
Essentially, we need to think of designs that move beyond task-based interventions and towards systems that sustain caregivers holistically over time.}

\edit{A major finding and contribution of our work is the model of evolution of the caregivers' mental health needs---an understanding of how their needs change across different stages of the caregiving journey. 
Caregivers move from initial uncertainty and emotional shock to long-term fatigue, grief, and burnout. 
Technologies that aim to support such evolving caregiver wellbeing must therefore be \textbf{designed to evolve in tandem with these changing emotional and informational needs}.}

\edit{Recent research in personal health informatics technology design highlights the importance of goal-aligned and adaptable systems that can support users' shifting motivations and priorities over time~\cite{munson2020importance,sefidgar2024migrainetracker}. 
\citet{munson2020importance} argued that successful health tracking tools should start with an understanding of users' personal goals and adjust as those goals evolve. 
Similarly,~\citet{sefidgar2024migrainetracker} showed how technologies that accommodate goal transitions (e.g., from symptom monitoring to emotional reflection) better align with the lived experiences of people managing chronic health conditions. 
Applying these insights into caregivers' mental wellbeing, we foresee a future where mental health platforms are adaptive. Tools that incorporate stage-sensitive mental health check-ins as well as adaptable self-care prompts and interventions that respond to caregivers' level of burnout, emotional triggers, or life transitions (for themselves and the care-recipients)---all with a goal of building good mental health skills and resilience.}

\subsection{Ethical Implications}

% \koustuv{Some caregivers showed optimism about AI and tech---however, what are the potential harms of AI?}

% \subsubsection{Data Security and Privacy}

As technology becomes more integrated into our lives and society, we also need to consider the ethical implications of adopting computing technologies in caregiving.
Some caregivers expressed concerns about the security of personal and sensitive information. 
In particular, some participants desired for more personalization, however, as ~\citeauthor{pandit2018ease} described, personalization can be a double-edged sword---personalization can come at the cost of more personal data, i.e., potentially compromised privacy.
This aligns with prior work on personalization-privacy tradeoffs~\cite{asthana2024know,li2024human,zargham2022want}.
Therefore, ensuring robust data encryption, secure storage, and compliance with regulations like the Health Insurance Portability and Accountability Act (HIPAA) is essential. 
Further, transparent privacy policies and giving users control over their data can enhance trust.
% \subsubsection{AI Reliability and Bias}

The recent emergence of generative AI and chatbots also brought up discussions about these AI chatbots in our interviews. 
In particular, the use of AI in caregiving tools raises questions about reliability and potential biases. 
AI systems may not fully comprehend the nuances of clinical conditions or cultural contexts, leading to inappropriate recommendations. 
Continuous evaluation and oversight of AI algorithms are necessary to prevent harm.
It is also important to incorporate caregivers' perspectives and needs in designing and building these chatbots---this will help utilize the values from lived experiences to anticipate AI harms and help in AI alignment~\cite{gabriel2020artificial,shen2024towards,saha2025ai,kawakami2023wellbeing,shi2025mapping}.
% \subsubsection{Equity and Access}
Further, there is a risk that technological solutions may widen the 
% gap of already existing 
digital inequalities~\cite{robinson2015digital}.
% between those who have access and those who do not. 
Efforts should be made to ensure that technologies are inclusive and accessible to caregivers of diverse backgrounds, including those with limited technological proficiency or resources.
% \subsubsection{Dependence on Technology}
Finally, we highlight the ethical dilemma of relying too heavily on technology in caregiving.
% also merits attention. 
While technology can supplement caregiving tasks, it cannot replace the empathy and emotional understanding that human interaction offers. 
Therefore, striking a balance between automation and human support is crucial to maintaining the dignity and emotional wellbeing of both caregivers and care recipients. 

% By proactively addressing these ethical implications, developers and policymakers can ensure that technological advancements benefit caregivers without compromising ethical standards.

%As we consider the role of technology in caregiving, several ethical implications arise. First, while AI tools offer promising support for caregivers, ethical responsibility to ensure accuracy and reliability is paramount. AI-driven platforms that offer medical or emotional advice must be held to high standards of evidence-based care to avoid potential harm from inaccurate recommendations.

%Furthermore, data privacy concerns are especially salient, as caregivers are often required to input sensitive information about themselves and their care recipients into digital platforms. Technologies that collect personal data must guarantee rigorous privacy protections and give caregivers full control over their data. Caregivers should have the right to opt out of data sharing, and developers should ensure that all data collection practices are transparent and secure.

%Lastly, the ethical dilemma of relying too heavily on technology in caregiving also merits attention. While technology can supplement caregiving tasks, it cannot replace the empathy and emotional understanding that human interaction offers. Striking a balance between automation and human support is crucial to maintaining the dignity and emotional wellbeing of both caregivers and care recipients.
 %2 page
% \vspace{-1.2em}
\section{Limitations and Future Directions}
Our study has limitations, which also point to interesting future directions.
The current study relied on a limited sample of U.S.-based caregivers and relied on participants' self-reported, which may not necessarily reflect the real-time AD/ADRD caregiving journey. 
% The participants in our study provided opinions of technology for their mental wellbeing. 
Based on the participants' perspective about technology for mental wellbeing, there is a clear need for future research to explore more potential uses and harms for technologies in mental wellbeing, and develop values elicitation technology that could help caregivers~\cite{foong2024designing}, especially in a collectivistic culture where there is an emphasis on harmonious decision-making~\cite{ozdemir2023caregiver, shin2013preferences}.
Future research can address these gaps by incorporating a more diverse sample across cultural contexts and conducting longitudinal studies that track the use of technology for mental health over time. Expanding research to develop and evaluate personalized, adaptive digital technology-including AI-driven and emotionally sensitive support systems -could provide more tailored mental health resources~\cite{wojcik2021informal} to meet the varying needs of caregivers more effectively. 
\edit{Future research could further explore how caregivers' mental health needs vary across different demographic groups and caregiving contexts. For instance, while financial burden emerged as a recurring theme in our study, we did not collect explicit data on participants' household income---a notable limitation. Examining the relationship between income levels and perceived financial strain in future work would offer valuable insights into the socioeconomic dimensions of caregiver wellbeing. }
Likewise, conducting participatory studies that involve other stakeholders, including community and institutional leaders, can enrich our understanding in building collaborative approaches and support strategies to cater to caregivers' mental wellbeing.
 %3 page
\section{Conclusion}
% \koustuv{conclusion needs to be updated with the new changes.}

Our study focused on the mental health of AD/ADRD caregivers, emphasizing the dynamic nature of their wellbeing needs and the role of technology in supporting their mental health during the caregiving journey. 
We conducted semi-structured interviews with 25 caregivers to analyze both the causes and effects of mental health concerns, \minor{developing a temporal mapping of how caregivers' mental wellbeing evolves across three distinct stages of the caregiving journey.}
% illustrating how mental wellbeing evolves over time. 
Our findings revealed that caregivers seek a combination of external support and self-care practices to address these concerns, highlighting the importance of accessible resources and coping strategies in mitigating mental health issues. 
Additionally, we explored caregivers' attitudes toward technology and the existing challenges they encounter, \minor{identifying the need for adaptive technologies that evolve alongside caregivers' challenging mental health needs rather than on-size-fits-all solutions.}
% highlighting opportunities for improvement. 
The proposed enhancements emphasize the need for accessible, personalized, and user-friendly technological support that addresses caregivers' unique and evolving needs. 
These insights contribute to designing more empathetic, caregiver-centered technological solutions and underscore the role of policy and community support for caregivers' mental health. 
By better understanding caregivers' multifaceted needs, our study provides a foundation for creating \minor{dynamic, stage-sensitive}
% tailored 
interventions that support caregivers holistically, benefiting both them and care recipient. 

 %3 page

%%
%% The acknowledgments section is defined using the "acks" environment
%% (and NOT an unnumbered section). This ensures the proper
%% identification of the section in the article metadata, and the
%% consistent spelling of the heading.

\begin{acks}
% \section{Acknowledgments}
This work was partly supported by the Jump ARCHES endowment through the Health Care Engineering Systems Center at Illinois and the OSF Foundation. Wang was supported by the University of Hong Kong's Overseas Research Fellowship (ORF) during the Summer of 2024. 
\end{acks}

%%
%% The next two lines define the bibliography style to be used, and
%% the bibliography file.
\bibliographystyle{ACM-Reference-Format}
\bibliography{0paper}

% \includepdf[pages=-,pagecommand={},width=1.2\textwidth]{CSCW Job Satisfaction Reviewer Responses.pdf}

%%
%% If your work has an appendix, this is the place to put it.
% \appendix
% \input{9appendix.tex}

\end{document}

\endinput
%%
%% End of file `sample-acmsmall.tex'.